\documentclass[5p,times,number]{elsarticle}
\pdfoutput=1
   
\usepackage{graphicx,miller,bm,booktabs,float,nicefrac,url,breakurl,color,microtype}
\usepackage{siunitx}[=v2]
\usepackage[version=4]{mhchem}
\usepackage{hyperref}
\usepackage[export]{adjustbox}

\hypersetup{breaklinks,hidelinks}

\DeclareMathAccent{\wtilde}{\mathord}{largesymbols}{"65}
\newcommand*{\matr}[1]{\bm{\mathit{#1}}}
\newcommand*{\tens}[1]{\bm{#1}}

\journal{Acta Materialia}

\begin{document}

\begin{frontmatter}

\title{Per-grain and neighbourhood stress interactions during deformation of a ferritic steel obtained using three-dimensional X-ray diffraction}

\author[a,b]{James A. D. Ball}
\author[c]{Anna Kareer}
\author[b]{Oxana V. Magdysyuk}
\author[b]{Stefan Michalik}
\author[b]{Thomas Connolley}
\author[a]{David M. Collins}
\ead{D.M.Collins@bham.ac.uk}
\affiliation[a]{organization={School of Metallurgy and Materials, University of Birmingham},
        addressline={Edgbaston}, 
        city={Birmingham},
        postcode={B15 2TT}, 
        country={United Kingdom}}
\affiliation[b]{organization={Diamond Light Source Ltd.},
        addressline={Harwell Science and Innovation Campus}, 
        city={Didcot},
        postcode={OX11 0DE}, 
        country={United Kingdom}}
\affiliation[c]{organization={Department of Materials, University of Oxford},
        addressline={Parks Road}, 
        city={Oxford},
        postcode={OX1 3PH}, 
        country={United Kingdom}}

\begin{abstract}
Three-dimensional X-ray diffraction (3DXRD) has been used to measure, in-situ, the evolution of \num{\sim 1800} grains in a single phase low carbon ferritic steel sample during uniaxial deformation.
The distribution of initial residual grain stresses in the material was observed to prevail as plasticity builds, though became less pronounced, and therefore less influential as strain increased.
The initial Schmid factor of a grain was found to be strongly correlated to the intergranular stress change and the range of stresses that are permissible; a grain well aligned for easy slip is more likely to exhibit a range of stresses than those orientated poorly for dislocation motion.
The orientation path of a grain, however, is not only dependent on its initial orientation, but hypothesised to be influenced by its stress state and the stress state of its grain environment.
A grain neighbourhood effect is observed: the Schmid factor of serial adjoining grains influences the stress state of a grain of interest, whereas parallel neighbours are much less influential.
This phenomenon is strongest at low plastic strains only, with the effect diminishing as plasticity builds.
The influence of initial residual stresses becomes less evident, and grains rotate to eliminate any orientation dependent load shedding.
The ability of the BCC ferrite to exhaust such neighbourhood interactions, which would otherwise be detrimental in crystal structures with lower symmetric and fewer slip systems, is considered key to the high ductility possessed by these materials. 
\end{abstract}


\end{frontmatter}

\section{Introduction}

To fully understand the response of a polycrystalline material to deformation, it is necessary to understand the factors that govern the onset of plastic deformation for individual grains.
The behaviour of individual grains must be influenced by any long-range stress acting over the whole sample or component, referred to as Type I stresses, but additionally from the presence of short/medium-range internal stresses at the microstructure scale---these are referred to as the Type II (grain average, intergranular) or Type III (intragranular, varying stress within a grain) stresses \cite{Withers2001}.
An important source of these inter- and intragranular stresses in a single phase material arises from dislocations and dislocation structures that form during plasticity \cite{MUGHRABI19831367,LEVINE20115803,KASSNER201344,ZHANG2022112113}.
For an idealised polycrystal, the resolved shear stress (RSS) is the degree of shear stress applied to a slip plane of an individual grain when an external load is applied \citep{taylor_plastic_1938}.
The RSS is a function of the applied stress and the orientation of the grain relative to the loading axis of the polycrystal.
When the RSS exceeds a critical value (the critical resolved shear stress or CRSS), for a slip plane within a grain, slip will initiate on that plane.
The RSS value presupposes, for an ideal polycrystal, that the onset of plastic strain within each grain is entirely governed by the orientation of the grain relative to the tensile axis.
Recent studies, however, suggest that even microstructurally simple alloys may deviate from these Taylor law predictions at the meso- and macroscale.
In austenitic stainless steels, \citet{juul_analysis_2020} found deviations from predictions in both grain lattice rotations and the dependence of stress state on grain orientation.
\citet{greeley_using_2019} found inconsistencies between measured and predicted resolved shear stresses in a simple \ce{Mg}-\ce{Nd} alloy; such studies reinforce the need for new experiments to reveal these fundamental aspects of deformation. 

The requirement for understanding the governing agents of deformation is certainly true of high-performance multiphase steels, possessing complex microstructures and processing methods; a complete understanding of the micro- and macromechanical mechanisms that govern plastic deformation is far from trivial.
However, there is evidence that even microstructurally simple steel alloys respond to deformation in unexpected ways.
Recent analyses of single phase ferritic steels have shown strain-path dependent deformation, in combination with initial texture, influences hardening \cite{collins_synchrotron_2015, collins_synchrotron_2017}.
The yield stress of interstitial-free steels has also been shown to depend on the distribution of dislocations, even with near-identical dislocation densities \citep{tanaka_effect_2020}.
This demonstrates an evolving understanding of even simple steel alloys and how they perform under deformation.

One factor that can modify the stress state (and therefore the onset of slip) for individual grains within a polycrystal is its immediate local neighbourhood.
\citet{kocks_texture_2000} found deviations in grain strain rates from Taylor model predictions, which were attributed to grain neighbourhood effects.
For a grain of interest, neighbourhood grains can generate stress concentrations that modify the stress state of the grain, depending on the orientation relationship between the grain and its neighbourhood grains.
Subsequent simulations by \citet{raabe_theory_2002} on the generation of orientation gradients within grains during loading found a significant dependence of orientation gradient strength on interactions between grains and their immediate neighbours.
Further simulation studies have noted grain neighbourhood interactions which significantly modify grain strain values \citep{bretin_neighborhood_2019}, leading to neighbourhood dependencies in dynamic strain aging \citep{gupta_crystal_2019}, void growth \citep{christodoulou_role_2021}, and fatigue life \citep{stopka_simulated_2022}.

Grain neighbourhood effects have also been directly observed experimentally. 
Using electron back-scatter diffraction (EBSD) and digital image correlation (DIC) methods, grain neighbourhood effects have been found to significantly influence deformation twinning in twinning-induced plasticity (TWIP) alloys \citep{gutierrez-urrutia_effect_2010}, strain localization in \ce{Ni}-based superalloys \citep{stinville_high_2015}, and grain boundary sliding in \ce{Al} alloys \citep{linne_effect_2020}.
These effects have also been studied in-situ using three-dimensional techniques.
Synchrotron X-ray studies have demonstrated a strong influence of grain neighbourhood on grain stress states in hexagonal close packed (HCP) polycrystals \citep{abdolvand_strong_2018}. 
These effects are partially attributed to the elastic and plastic anisotropies inherent to HCP systems. 
However, grain neighbourhood effects may have been observed in-situ in cubic systems by \citet{neding_formation_2021}, who found stacking faults being generated in grains that were poorly orientated for fault formation according to their Schmid factor.

Another significant parameter that influences the response of a grain to an external load is the residual stress of the grain. 
These residual stresses can be desirable---residual compressive stresses induced by shot peening, for example, can significantly improve resistance to fatigue crack initiation \citep{trung_effect_2017}.
However, they can be deleterious to material performance: for example by driving creep cracking at elevated temperatures in stainless steels \citep{turski_residual_2008}.
Residual stresses can be generated by many material processing operations, such as welding \citep{withers_residual_2002, deng_numerical_2006}, forging \citep{atienza_residual_2005} and casting \citep{liu_study_2001}.
Significant efforts have been made to simulate the formation of residual stresses during such operations \citep{deng_numerical_2006, liu_study_2001, rong_review_2018}, and to explore the effects of initial residual stresses on further deformation \citep{withers_recent_2008, turski_residual_2008}.
Although direct observation of residual stresses in three-dimensional components is non-trivial, synchrotron radiation has been used to quantify them in two-dimensional projections \citep{withers_residual_2002, withers_recent_2008}.

It is evident that direct in-situ measurements of individual grain responses to deformation are required to further explore these effects. 
Far-field Three-Dimensional X-Ray Diffraction (ff-3DXRD) is an ideal technique for measuring in-situ deformation response at the mesoscale.
Using ff-3DXRD, per-grain centre-of-mass positions, orientations, and Type II strain states can be evaluated for many hundreds of grains simultaneously in large samples \citep{oddershede_determining_2010}, and recent advancements to the technique can also measure Type III stresses and grain morphologies \citep{henningsson_intragranular_2021}. 
In the extreme case, with far-field 3DXRD alone, highly accurate characterisation of grain neighbourhoods can be performed in-situ for thousands of grains \citep{louca_accurate_2021}. 
These measurement capabilities have enabled newfound microstructural insights into the deformation response of simple steel alloys \citep{offerman_following_2012, guillen_situ_2018, juul_analysis_2020}. 
These techniques can also be utilised for the analysis of more complicated alloys such as shape-memory alloys \citep{el_hachi_multi-scale_2022}, duplex stainless steels \citep{hedstrom_load_2010} and transformation-induced plasticity (TRIP) steels \citep{jimenez-melero_effect_2009}.
Of particular note is recent work by \citet{el_hachi_multi-scale_2022}, who used in-situ 3DXRD to observe grain-neighbourhood effects deformation-induced martensitic transformations (DIMT) in a \ce{Cu}-\ce{Al}-\ce{Be} alloy. 

In-situ 3DXRD experiments are currently limited to specialised synchrotron beamlines around the world and are therefore not commonplace. 
New instruments that enable reliable collection and analysis of 3DXRD data are highly desired by the materials science community. 
The aim of this study is two-fold: to establish and evaluate in-situ far-field 3DXRD at the I12 beamline at Diamond Light Source, and to use the technique to explore the effect of residual stress and grain neighbourhood on grain responses to in-situ deformation in 3D.

\section{Experimental Method}
\subsection{Material}

A tensile dog-bone sample with a \SI{1}{\milli\metre} wide gauge section and a \SI{6}{\milli\metre} gauge length was produced from a \SI{1}{\milli\metre} thick Zn-galvanized sheet of DX54 steel, a single-phase ferritic steel with a body-centered cubic crystal structure.
The nominal composition is given in Table \ref{tab:DX54_composition}.
Prior to any testing, the galvanized surface was removed with abrasive media, then annealed at \SI{980}{\celsius} for \SI{1}{\hour} and slow-cooled at \SI{\sim 1}{\celsius\per\minute} to achieve a coarse equiaxed microstructure.

\begin{table}
    \caption{Nominal chemical composition of DX54 steel \cite{collins_synchrotron_2015}.}
    \begin{tabular}{@{}l|ccccc@{}}
        Element & Fe & C & P & S & Mn\\
        wt.\% & Balance & $\leq 0.06$ &$ \leq 0.025$ & $\leq 0.025$ & $\leq 0.35$
    \end{tabular}

    \label{tab:DX54_composition}
\end{table}

\begin{figure}
    \includegraphics[width=\columnwidth]{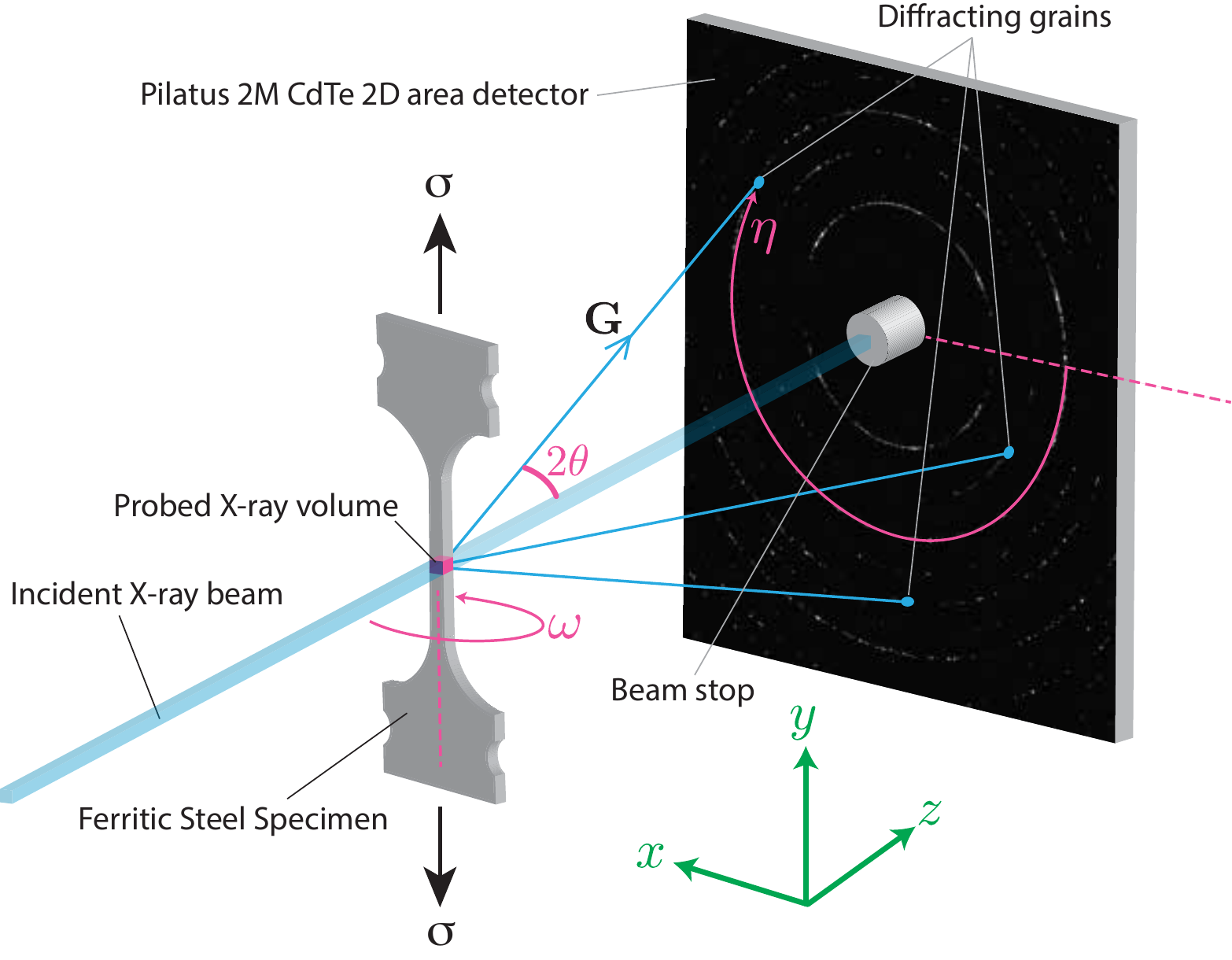}
    \caption{Tensile specimen and 3DXRD configuration at the I12 beamline, Diamond Light Source.}
    \label{fig:expt_diagram}
\end{figure}

\subsection{3DXRD Data Acquisition}

3DXRD data were collected in Experimental Hutch 1 of the I12 beamline at Diamond Light Source \cite{drakopoulos_i12_2015}.
The experimental geometry is shown in Figure~\ref{fig:expt_diagram}; the axes represent the sample reference frame that will be hereon used.
Prior to loading a sample, a multi-distance calibration \cite{hart_complete_2013}, executed within DAWN \cite{basham_data_2015,filik_processing_2017} was performed with a NIST 674b \ce{CeO2} standard reference sample \cite{cline_powder_2016}.
The energy of the monochromatic X-ray beam was determined as \SI{60.2}{\kilo\electronvolt} and the sample to detector distance was \SI{550.3}{\milli\metre}.
The tensile specimen was placed in a Deben CT5000 5~kN load frame designed for X-ray tomography scans, with the loading axis set to be axisymmetric with the sample stage rotation axis. 
3DXRD data were collected at multiple applied vertical ($y$-axis) loads to allow tracking of individual grains under different applied external loads.

Ten successive ``letterbox" 3DXRD scans were taken along the sample gauge ($y$ direction) with a \SI{1.5 x 0.15}{\milli\metre} beam, and \SI{0.05}{\milli\metre} of overlap between each.
The total illuminated sample volume at each load step was \SI{\sim 1 x 1 x 1}{\milli\metre}.
During each letterbox scan, the sample was rotated about the $y$ axis from \SI{-180}{\degree} to \SI{180}{\degree}.
Diffraction patterns were acquired with a Pilatus 2M \ce{CdTe} area detector, recording data at \SI{1}{\degree} increments and a \SI{1}{\second} exposure time.
This data collection procedure replicated a previously established routine for the beamline \cite{ball_implementing_2022}.

\begin{figure}[h]
    \centering
    \includegraphics{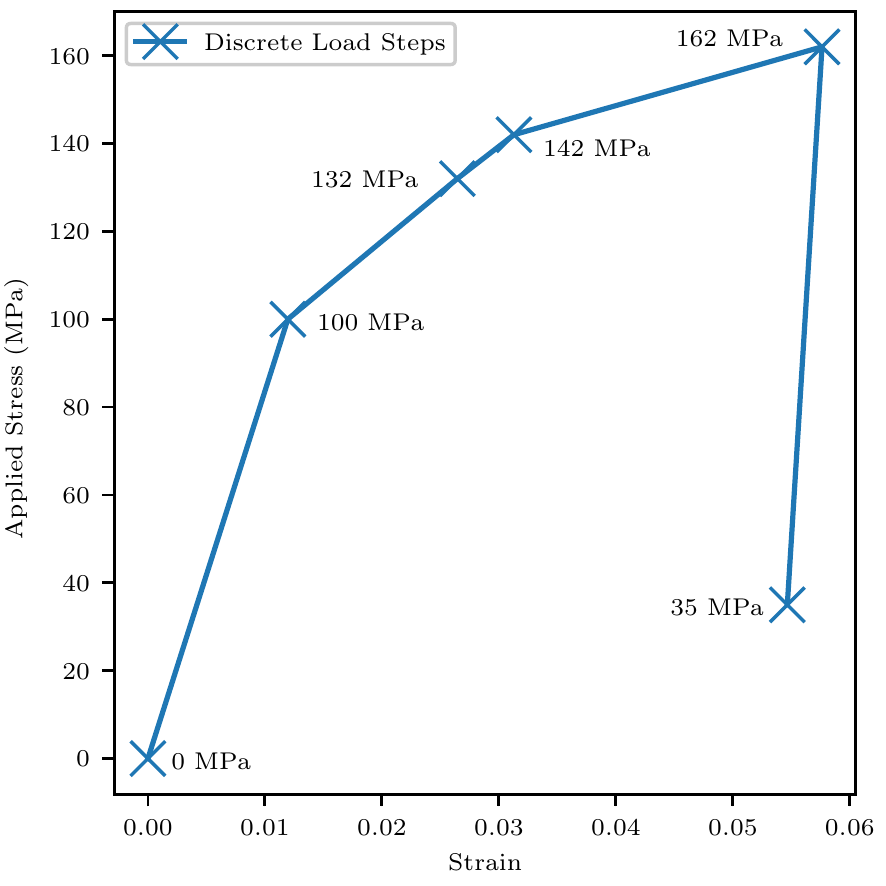}
    \caption{Stress-strain increments at which 3DXRD measurements were made.}
    \label{fig:load_steps}
\end{figure}

The load frame was operated in displacement control at a constant rate of \SI{0.2}{\milli\metre\per\minute} until the desired force targets were reached.
At each deformation step, the applied load was held constant.
Six load steps, shown in Figure~\ref{fig:load_steps}, were chosen to explore the deformation response up to a small degree of plasticity (maximum strain of \SI{\sim 5}{\percent}).
The final load step was recorded after unloading the sample as far as possible, to investigate the remaining residual stresses in the sample.
Two small strips of Kapton tape were affixed at each end of the sample gauge area to act as fiducial markers. 
The macroscopic strain on the sample was measured from large field-of-view radiographs, acquired with an end-of-hutch X-ray imaging camera, at each load step using the fiducial marker separation.

\subsection{EBSD data collection}

For EBSD data collection, the sample was polished to a \SI{0.04}{\micro\metre} surface finish using colloidal silica, then examined with a Zeiss Merlin field emission gun scanning electron microscope (FEG-SEM).
A Bruker $\text{e}^{-}\text{Flash}^{\text{HR}}$ EBSD detector was used to collect EBSD maps at a \SI{5}{\nano\ampere} probe current and a \SI{20}{\kilo\electronvolt} beam energy.
Electron backscatter diffraction patterns (EBSPs) were recorded at high angular resolution (HR-EBSD), \num{800 x 600} pixels$^2$ and were saved for post-collection analysis; the scan and indexing was performed using Esprit 2.0 software.
A low magnification map (\SI{5 x 6.5}{\milli\metre}, \SI{2}{\micro\metre} step size) of the sample was also taken in the undeformed grip region.
A higher spatial resolution map (\SI{1.0 x 0.8}{\milli\metre}, \SI{2}{\micro\metre} step size) of the deformed specimen, within the gauge of the tensile specimen was also performed.

HR-EBSD was used to estimate intragranular residual elastic Type III stresses from the deformed specimen using an in-house written method that measures subtle changes in the crystal geometry, inferred from the EBSPs.
This method extracts a reference EBSP within a grain, from which image shifts are measured via a cross correlation function between the reference pattern and test patterns within this grain.
A deformation gradient tensor is defined from the diffraction pattern shifts, from which strain and rotation components can be separated using a finite decomposition framework \cite{britton_2018}.
The method has a strain sensitivity of at least \num{1e-4} \cite{villert2009}.
The resulting strain tensor, for each pixel, is multiplied by the stiffness tensor for ferritic steel, using the Voigt notation, to obtain the stress tensor.
Comprehensive details of the HR-EBSD method are provided elsewhere \cite{WILKINSON2012366,Britton2013}, as well as the mathematical descriptions \cite{WMG2006_1,WILKINSON2006307}. 

\subsection{3DXRD analysis}

\begin{figure}
    \centering
    \includegraphics[width=\columnwidth]{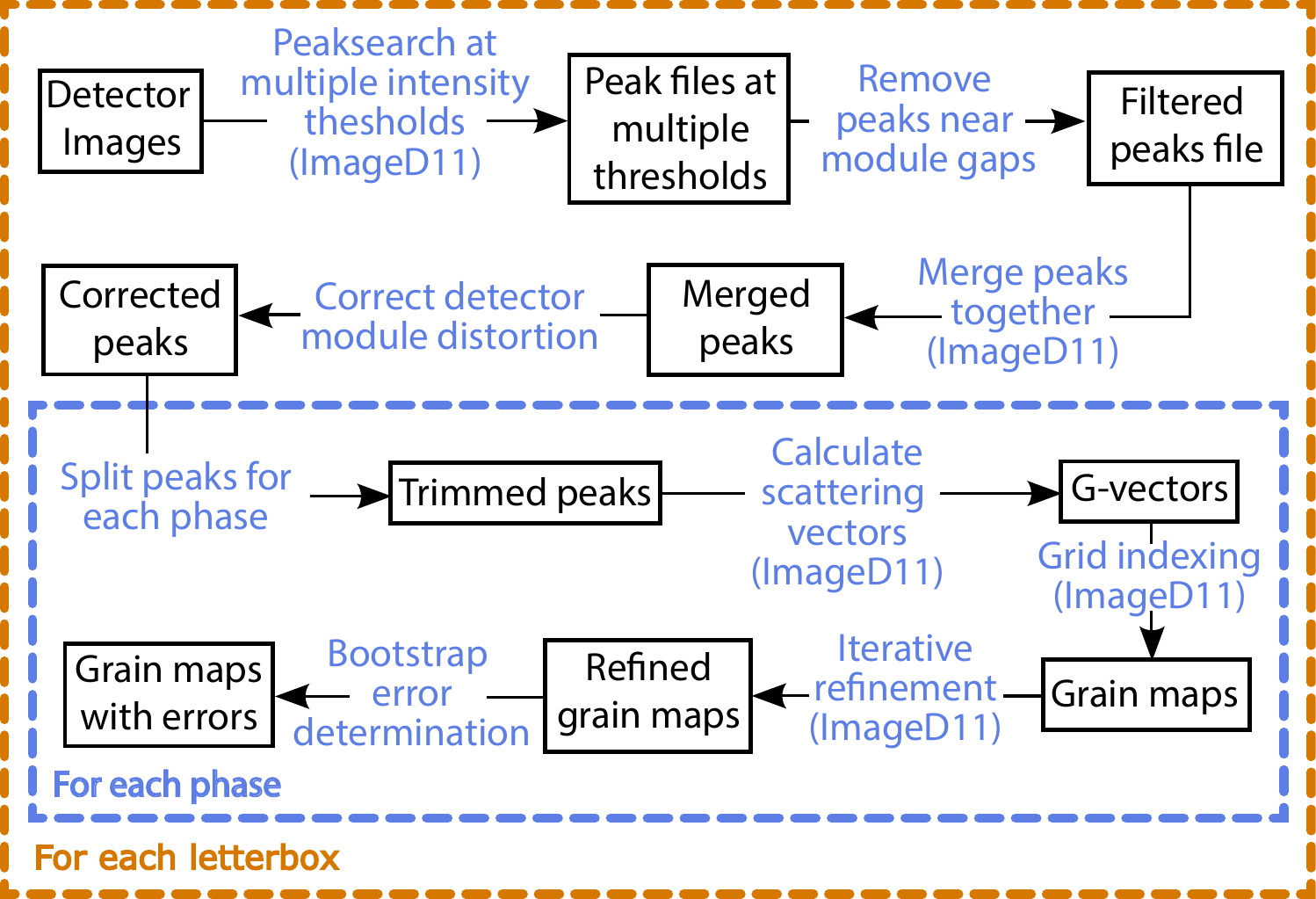}
    \caption{Indexing and analysis procedure for 3DXRD datasets.}
    \label{fig:flowchart}
\end{figure}

Indexing and analysis of collected 3DXRD data broadly followed routines established in \cite{ball_implementing_2022}. 
However, a number of improvements have since been made to the analysis procedure that yield significant gains in data quality. 
A flowchart showing the data processing steps utilised in this study are shown in Figure~\ref{fig:flowchart}.
A number of pre-processing stages have been introduced to reduce error in detector peak positions. 
Peaks close to gaps in the modules of the Pilatus diffraction detector were deemed unreliable due to inaccurate intensity profile shape and were removed.
As the position and orientation for each module on the detector is independent, small distortions in peak position may have been introduced if these displacements were not corrected for. 
Following an established routine \cite{wrightUsingPowderDiffraction2022} applied to a series of reference \ce{CeO2} calibration patterns, the module displacements were determined---this allowed a correction file to be generated which specifies the sub-pixel adjustments required to correct the peaks data.
Diffraction peak searching and determination of grain positions, orientations and strain states was performed with ImageD11 \cite{fable}, with each individual letterbox scan indexed separately.

To enable higher-level analyses of the 3DXRD data, as a complete dataset, in-house pre- and post-processing software was developed using the Python programming language.
This software coordinated the parallel initial indexing of each ImageD11 letterbox scan. 
An object-orientated model was devised to store and manipulate grain data at a number of different levels: the sample as a whole; individual load steps; individual scans within those load steps; single phases within those scans; and finally individual grains within each phase.
This data model enabled sophisticated post-processing and data analysis, such as duplicate grain detection, sample rigid body transform detection, grain nearest neighbour identification, and positional grain filtering.

A new ``bootstrap" approach was devised to determine errors in grain positions, orientations, and strains. More specifically, the grain Biot strain and stress tensors were determined from the reciprocal lattice lengths of the grain as described in \ref{sec:tensors}.
After each grain was indexed with the grid indexing procedure, the detector peaks associated with that grain were isolated.
Next, 100 copies of these peaks were generated.
For each copy, a random \SI{50}{\percent} of the peaks were removed. 
Then, using each copy containing only \SI{50}{\percent} of the peaks, the grain position, orientation and strain was refined using ImageD11. 
This allowed the convergence of the grain parameter refinement to be probed---if a large variation in grain parameter outcomes was observed, it would indicate a high degree of error in each peak, leading to a poor degree of convergence of the refinement routine. 
Conversely, if only a small variation in grain parameter was observed between refinements, the error in each peak would have to be lower. 
The means of the grain positions, orientations and strain tensor elements of each copy were taken as the final parameters for that grain. 
The standard deviations of the parameter distributions were used as the errors for those grain parameters.

After the bootstrap routine, a cleanup process removed any duplicated grains created by the grid index procedure. 
Using these cleaned grain data, each individual ``letterbox" scan was stitched together to form a single grain map for each load step, with duplicate grains removed in the overlap regions. 
Grains were then tracked across subsequent load steps to identify grains common to all load steps. 
For the cleanup, stitching and tracking stages, the same grain de-duplication algorithm, described in detail in \ref{sec:dedup}, was used to identify and group together repeated grain observations.

Once grains were tracked across multiple load steps, tracked grain positional data was used as an input to a rigid body transformation solver using the coherent point drift algorithm as implemented in the pycpd Python library \cite{gatti_pycpd_2022}. 
The resultant orientation matrix was then used to generate a modified orientation matrix at a load-step level in which individual grains could be subsequently assessed, with any sample rotations effectively removed.

\begin{figure}[h!]
\centering
    \includegraphics{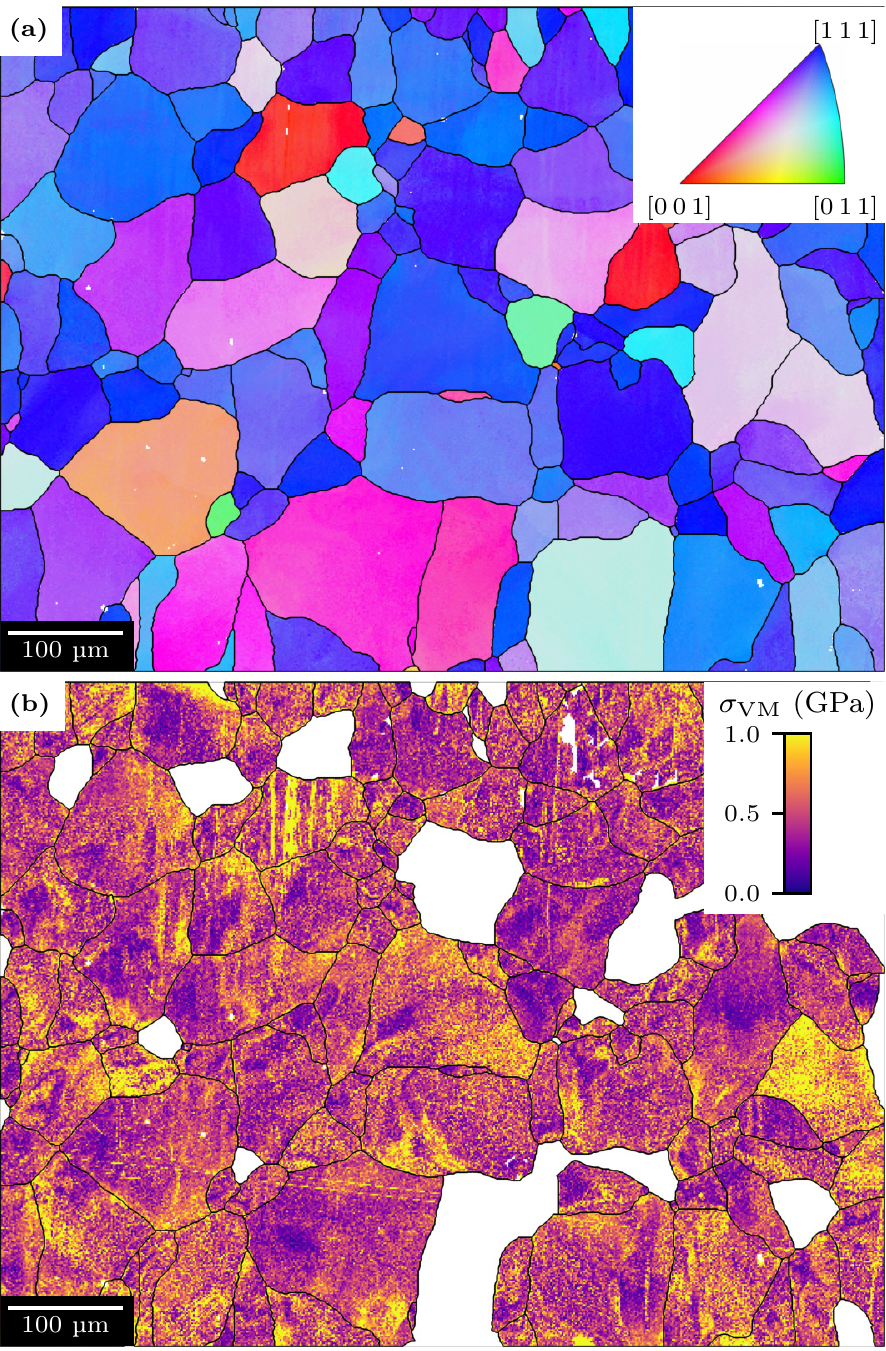}
    \caption{HR-EBSD map obtained from the sample gauge, post deformation with \SI{\sim 5}{\percent} plastic strain; (a) IPF-$Z$ colouring and (b) the corresponding per-pixel Type III von~Mises stresses. The white regions are grains determined to have a low quality and were eliminated from the analysis. The loading axis is left-right.}
    \label{fig:hr_ebsd}
\end{figure}

\newpage

\section{Results}

\begin{figure}[h!]
    \centering
    \includegraphics{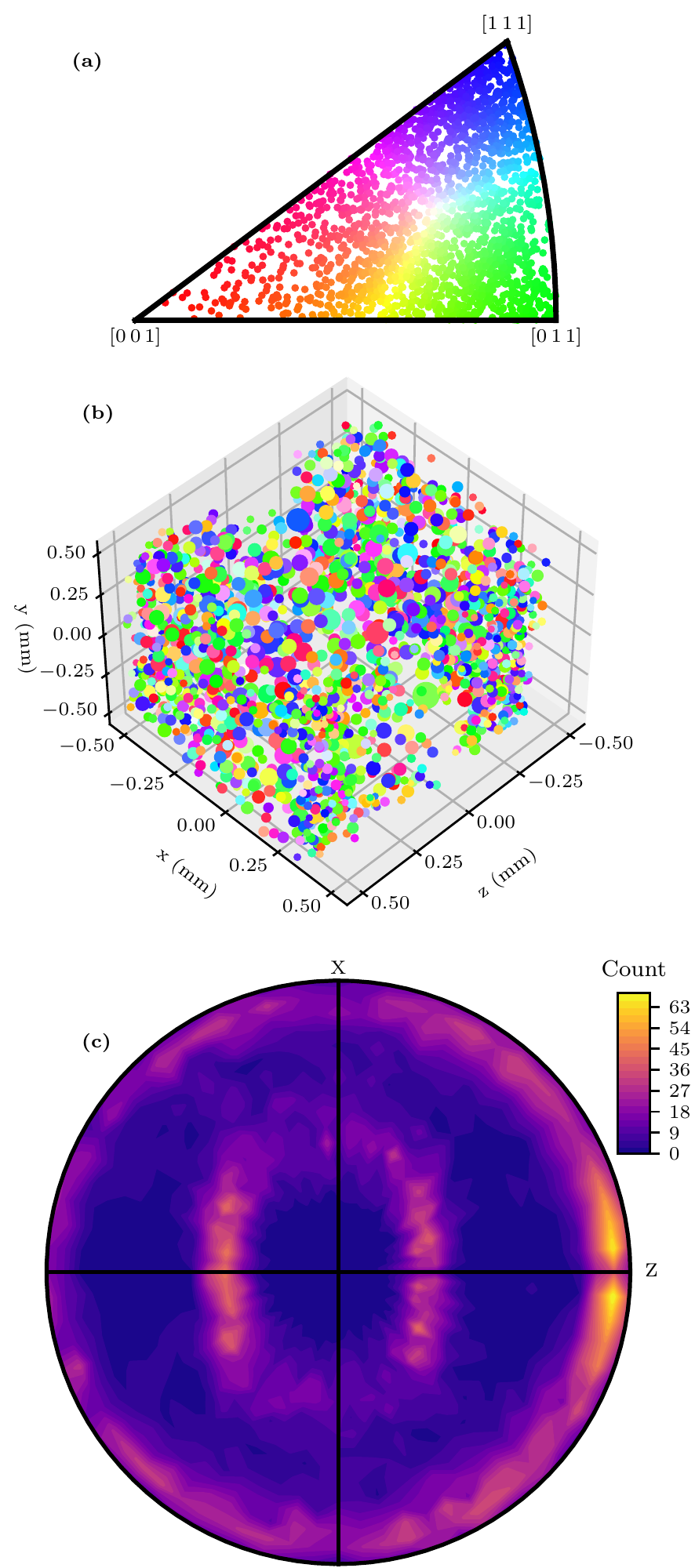}
    \caption{In the condition prior to loading, grains within the tensile specimen gauge region have crystal orientations shown on $Z$-axis inverse pole-figure (a), with the corresponding colours shown in (b) grain position map, and (c) a texture evident on the $Z$-axis direct \hkl{110} contoured pole-figure.}
    \label{fig:no_load_map_orien}
\end{figure}

\subsection{Microstructure}

The grain mean spherical equivalent diameter was determined to be \SI{130}{\micro\metre}, calculated from a low-magnification EBSD scan of \num{1797} grains with a \SI{2}{\degree} misorientation tolerance to determine the grain boundaries.
The reader is referred to Figure~1 in the Supplementary Materials for an IPF-$Z$ EBSD map of these results.
The sample examined with 3DXRD was also characterised post-deformation with a HR-EBSD scan acquired within the sample gauge.
An IPF-$Z$ map is shown in Figure~\ref{fig:hr_ebsd}a.
Using the stress (residual elastic Type III) tensor elements obtained from the cross-correlation method, the per-pixel von~Mises stress, $\sigma_{\rm VM}$, was calculated, shown in Figure~\ref{fig:hr_ebsd}b.
The reader is referred to Figure~3 in the Supplementary Materials for maps of each stress tensor component of the HR-EBSD scan.
The magnitude of these localised intraganular stresses is seen to far exceed the macroscopic yield stress of the material (\SI{\sim 100}{\mega\pascal}), with evidence of steep stress gradients in most grains.
Stress banding is also evident in several grains, featuring as yellow, approximately vertical streaks that traverse part or whole grains; these structures are evidence of residual stresses developing from plasticity structures. 

\subsection{3DXRD grain indexing}

In total, \num{18320} raw grains were indexed over the six load steps. 
After the grain de-duplication and letterbox stitching routines, \num{10741} grains remained, with an average of \num{1790} grains per load step.
Table~\ref{tab:grain_numbers} shows the number of stitched grains remaining at each load step.
\begin{table}[H]
\centering
\caption{Number of stitched grains remaining per load step.}
\label{tab:grain_numbers}
\begin{tabular}{@{}SS@{}}
\toprule
\multicolumn{1}{c}{$\sigma_{\rm{Applied}}$ (\si{\mega\pascal})} & \multicolumn{1}{c}{Number of stitched grains} \\ \midrule
0                  & 2012                      \\
100                & 2343                      \\
132                & 1999                      \\
142                & 1817                      \\
162                & 1275                      \\
35 {(}unload{)}        & 1295                  \\ \bottomrule
\end{tabular}
\end{table}

An overview of the sample volume probed, given for an example unloaded state (prior to deformation) is shown in Figure~\ref{fig:no_load_map_orien}.
The distribution of the grain orientations are represented on an inverse pole figure, and the corresponding grain position map is shown in Figure~\ref{fig:no_load_map_orien}a and Figure~\ref{fig:no_load_map_orien}b, respectively.
Here, the grains are coloured by their orientation and the size of each point in the grain position map is scaled by the grain volume.
The centre-of-mass positions of indexed grains is seen to conform well to the sample geometry, representing the probed volume of the tensile specimen within the gauge.
A rolled texture is observed in grain orientations as shown in the \hkl{110} pole figure; Figure~\ref{fig:no_load_map_orien}c.

As the original DX54 sample sheet was hot-dip galvanized with a \ce{Zn} coating, a variation in lattice parameter was expected in grains located close to the original galvanised surfaces.
This arises from any remnant surface Zn diffusing into the surface during the heat treatment stage used to tailor the grain size.
To investigate this, grain unit cells were extracted, as shown in Figure~2 in the Supplementary Materials. 
For grains close to the surfaces where a variation in lattice parameter is evident, a geometric filter was applied to the 3DXRD dataset to remove grains with centre-of-mass positions more than \SI{0.3}{\milli\metre} from the origin along $x$ or $z$.
The remaining grains can then be safely evaluated for changes in strain.

\subsection{Stress development}

\begin{figure}[h]
    \includegraphics{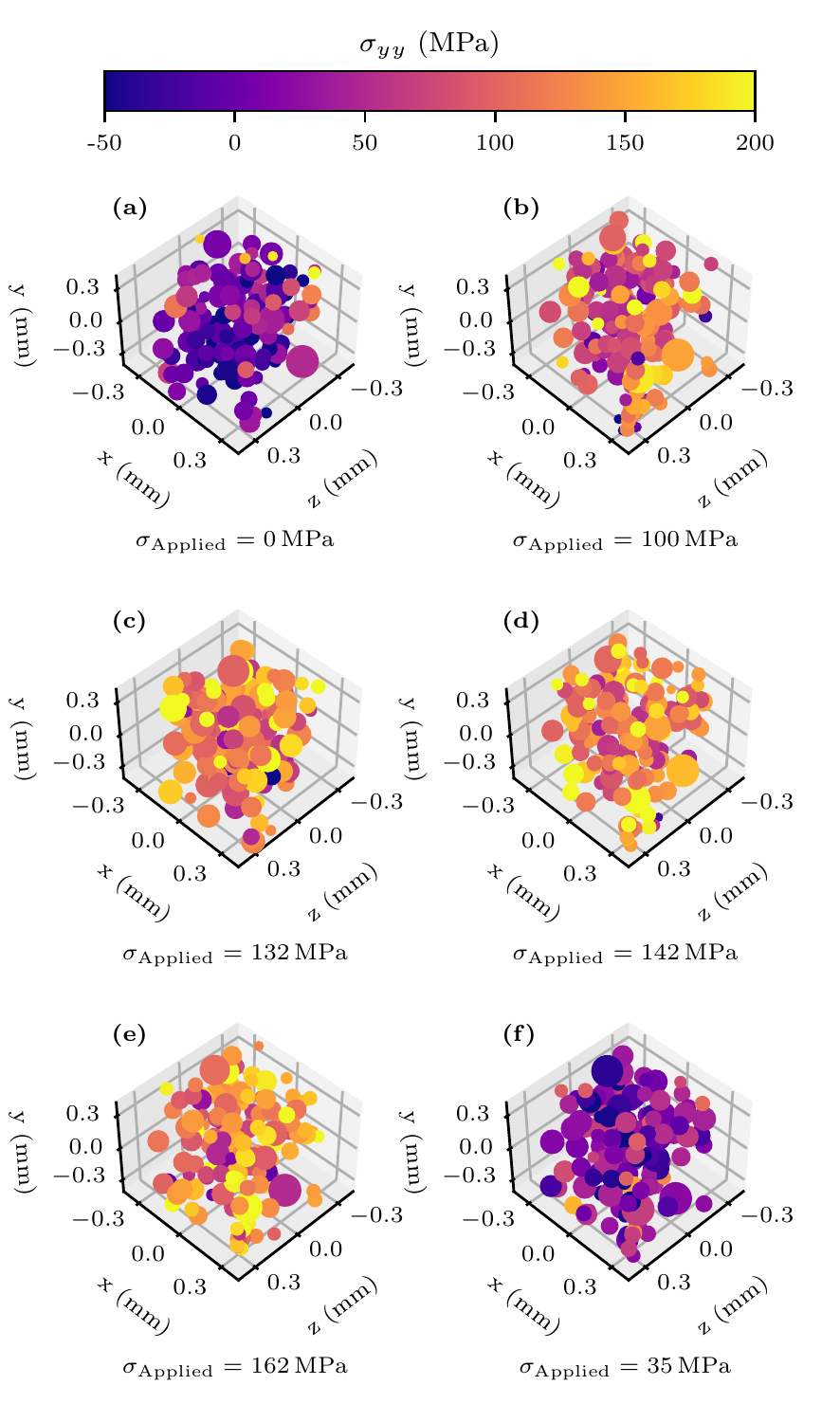}
     \caption{Grain maps at multiple applied loads, coloured by the vertical, $\sigma_{yy}$, stress in the sample reference frame (tensile direction).}
    \label{fig:s33_map_grid}
\end{figure}

\begin{figure*}[h!]
    \includegraphics[width=190mm,center]{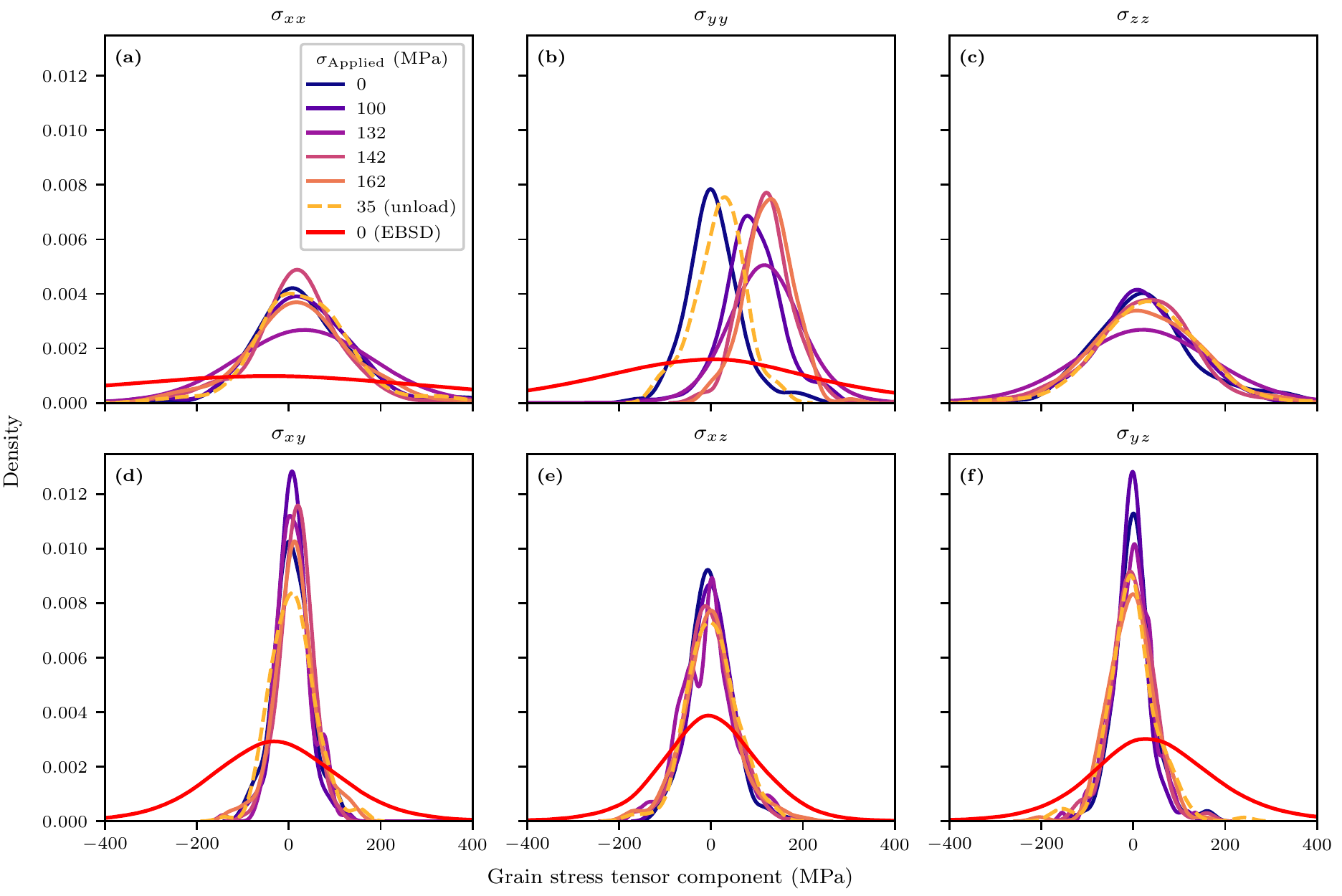}
    \caption{3DXRD Type II grain stress distributions across multiple load steps, with added EBSD Type III grain stress distributions.}
    \label{fig:stress_distributions}
\end{figure*}

With each indexed grain assigned a stress tensor, position, orientation and size for every load step, general trends for the deformation accumulation in the ferritic steel can be described. 
Figure~\ref{fig:s33_map_grid} shows the filtered grain centre-of-mass map at multiple load steps, coloured by the vertical component of the stress vector in the lab frame (co-axial with the applied load from the load frame). 
The changes in grain stress over the loading sequence are clearly visible; the mean average stresses follow the applied stress steps, with a wide distribution.
Notably, a wide range in initial grain stresses was observed in the first load step ($\sigma_{\rm{Applied}}=$~\SI{0}{\mega\pascal}), although this becomes less pronounced after unloading ($\sigma_{\rm{Applied}}=$~\SI{35}{\mega\pascal}).
To investigate the grain stress distributions further, histograms are plotted for each component of the 3DXRD Type II stress tensor, $\tens{\sigma}$, in the lab frame, as shown in Figure~\ref{fig:stress_distributions}.
The components of the HR-EBSD Type III stress tensor from the post-deformation condition have also been plotted.
At no applied load, a broad distribution in $\sigma_{yy}$, centered roughly at the origin, is observed.
This distribution narrows significantly and shifts positively as strain increases in the sample, indicating that the substantial initial residual stresses play a less influential role as plasticity builds.
The distributions in the transverse normal stresses ($\sigma_{xx}$ and $\sigma_{zz}$) are also initially broad, but the degree of narrowing is reduced compared to $\sigma_{yy}$.
The shear components, $\sigma_{xy}$, $\sigma_{xz}$ \& $\sigma_{yz}$ show a stress distribution that is characteristically narrower than the linear components.
As stress is applied, these distributions broaden slightly, with bumps developing in the leading and trailing tails.
This demonstrates several grains that differ from the majority as plasticity builds.
In every direction, the Type III grain stress distributions from the HR-EBSD dataset show substantially broader distributions than the 3DXRD Type II stresses, indicating that the localised grain stresses can deviate significantly, higher or lower, than the grain averaged Type II stress as measured by 3DXRD.
Additionally, the $\sigma_{xy}$ shear component in the EBSD scan is narrower than the axial stress component ($\sigma_{xx}$), which is narrower still than the transverse stress component ($\sigma_{yy}$), again mirroring the behaviour of the Type II grain stresses.

\subsection{Grain tracking}

\begin{figure*}[h]
    \centering
    \includegraphics{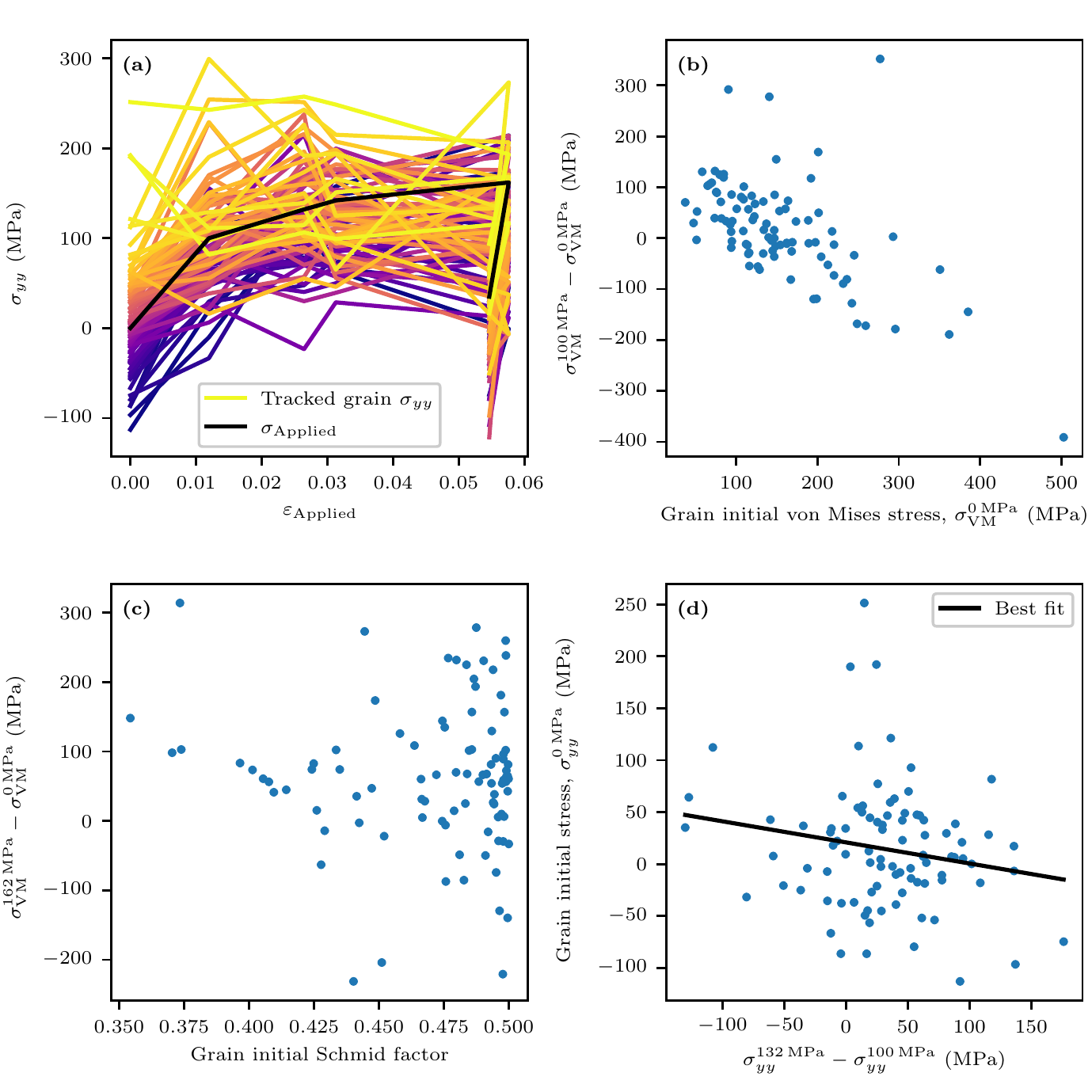}
    \caption{Tracked grain correlations: (a): following macroscopic yield, coloured by their initial $\sigma_{yy}$ values; (b): initial von~Mises stress vs change in von~Mises stress between first two load steps ($\sigma_{\rm{Applied}} = \SI{0}{\mega\pascal}$ and $\SI{100}{\mega\pascal}$); (c): initial Schmid factor vs change in von~Mises stress between no load and max applied load ($\sigma_{\rm{Applied}} = \SI{0}{\mega\pascal}$ and $\SI{162}{\mega\pascal}$); (d): $\sigma_{yy}$ change between $\sigma_{\rm{Applied}} = \SI{132}{\mega\pascal}$ and $\SI{100}{\mega\pascal}$ vs initial $\sigma_{yy}$.}
    \label{fig:combo_graph}
\end{figure*}

After the individual letterbox scans were stitched together, the stitched grain maps were tracked over all applied load steps, yielding \num{674} fully-tracked grains.
\num{178} fully-tracked grains remained after geometric filtration.
The evolution of grain parameters can therefore be explored across multiple load steps.
Figure~\ref{fig:combo_graph}a depicts the development of $\sigma_{yy}$ stress (tensile direction) in the lab frame for each tracked grain.
In general, good agreement is observed between the external applied load from the load frame and the individual tracked grain data. 
Most grains follow the trend of the macroscopic stress-strain curve. 
Grains with high initial stresses (residual stress) tend to maintain a higher stress state throughout the loading series. 
There are several grains that experience a stress drop when passing the yield point, a phenomenon also observed by \citet{hedstrom_load_2010} (in a duplex steel) and \citet{abdolvand_strong_2018} (in \ce{Zr} and \ce{Ti}).

The influence of grain initial (residual) stress on the change in grain stress between the first two load steps is examined in Figure~\ref{fig:combo_graph}b, where a clear negative correlation is observed between the initial total stress of a grain and its ability to further increase in total stress.
The relationship here indicates a grain with a low initial stress may experience a significant increase in von~Mises stress, whereas a grain with a high initial stress is more likely to see only a small increase, or a stress drop. 

The influence of the initial grain orientation on the grain stress development can be analysed, as per Figure~\ref{fig:combo_graph}c.
Grains with a lower initial Schmid factor are much more likely to increase rather than decrease their von~Mises stress over the loading series.
Grains with a higher Schmid factor have a wider range of permissible von~Mises stress changes over the loading series; the stress of these grains may increase or decrease.
This provides strong evidence that the orientation alone does not determine whether a grain hardens or softens, at the grain average stress level.

Figure~\ref{fig:combo_graph}d shows that the extent of the stress drop (after plastic deformation has commenced) appears to be negatively correlated with grain residual stress ($\sigma_{\rm{Applied}} = \SI{0}{\mega\pascal}$).
Grains with a higher residual stress tended to experience a greater stress drop.
As proposed by \citet{hedstrom_load_2010}, this may be evidence of the activation of multiple competing slip processes beyond a certain threshold strain that act to lower the overall lattice strain of the grain.

\section{Discussion}

During this investigation, the deformation behaviour of a single phase ferritic steel has been studied to reveal the interplay between the size, orientation, position and elastic stress/strain state, on a per-grain basis, within a volume of interest.
The interplay between these parameters is considered key to understanding the macroscopic behaviour of engineering alloys, where the method of far-field 3DXRD has been utilised for this purpose.
Following an initial proof of concept on the I12 beamline \citep{ball_implementing_2022}, the method implementation and its analysis have been developed significantly, as part of this work, enabling a first in-situ 3DXRD investigation at Diamond Light Source.
This discussion includes firstly a critique of the data processing and analysis method created for Diamond, followed by the resulting phenomenological micromechanical mechanisms that govern and control the tensile response of the ferritic steel during the onset and low levels of plasticity.
The addition of complementary HR-EBSD orientation and Type III stress measurements to supplement the per-grain 3DXRD observations are also discussed.

\subsection{Indexing quality}

An average of \num{1790} grains per load step were indexed, after grain de-duplication and letterbox stitching, with \num{2012} grains remaining in the first load step.
The success of the indexing strategy utilised here is evident in the grain map (Figure~\ref{fig:no_load_map_orien}); the edges of the sample are well defined, matching the macroscopic dimensions of the sample.
Any spurious ``satellite" grains in unfeasible positions relative to the sample are absent, indicating satisfactory convergence in the grain centre-of-mass position refinement.
The measured texture also matches other prior studies from DX54 steel of similar pedigree \cite{collins_synchrotron_2015, collins_synchrotron_2017,ball_implementing_2022}. 

\begin{table}[H]
\centering
\caption{Indexing technique precision summary for 3DXRD datasets obtained from I12, Diamond.}
\label{tab:error_comparison}
\begin{tabular}{@{}lc@{}}
\toprule
\begin{tabular}[c]{@{}l@{}}Parameter\\ \end{tabular}  & \begin{tabular}[c]{@{}c@{}}Error\\ \end{tabular} \\ \midrule
Orientation (\si{\degree}) & 0.03 \\
Position, horizontal (\si{\micro\metre}) & 8 \\
Position, vertical (\si{\micro\metre}) & 6 \\
$\varepsilon_{xx}$ ($\times \num{e-3}$) & 0.2 \\
$\varepsilon_{yy}$ ($\times \num{e-3}$) & 0.1 \\
$\varepsilon_{zz}$ ($\times \num{e-3}$) & 0.2 \\ \bottomrule
\end{tabular}
\end{table}
A summary of the grain parameter precision is given in Table~\ref{tab:error_comparison}.
The quoted values are derived from the parameter distributions, produced from the bootstrap data analysis; they represent one standard deviation from each distribution.
The magnitudes are comparable to those obtained by other authors \cite{Bernier2011}.
The orientation precision achieved here is attributed to the detector module distortion corrections; without this inclusion the orientation error is significantly higher (\SI{0.1}{\degree} \cite{ball_implementing_2022}). 
The position and strain accuracy are determined by the indexing and refinement procedure, as well as grain errors calculations. 
The high precision is evidence of a low degree of divergence between individual refinements during the bootstrap error determination process---this is due to the high number of average peaks per grain, and an accurate peak location afforded by the detector module distortion corrections. 
The grain strain error of 3DXRD at I12 at Diamond is considered acceptable in the context of engineering alloys, and is approximately one decade from state-of-the art implementations of the method, such as \num{1e-5} at the ID11 beamline of the ESRF \citep{oddershede_determining_2010}.
Further improvements to grain strain accuracy with the current detector are only likely to be possible by reducing the $\omega$ step size in subsequent experiments.
Using the present data analysis method, the precision in orientation and grain centre-of-mass position are now limited by the experimental geometry, detector pixel size, and detector dynamic range.

\subsection{Errors}

\begin{figure}[t!]
    \includegraphics{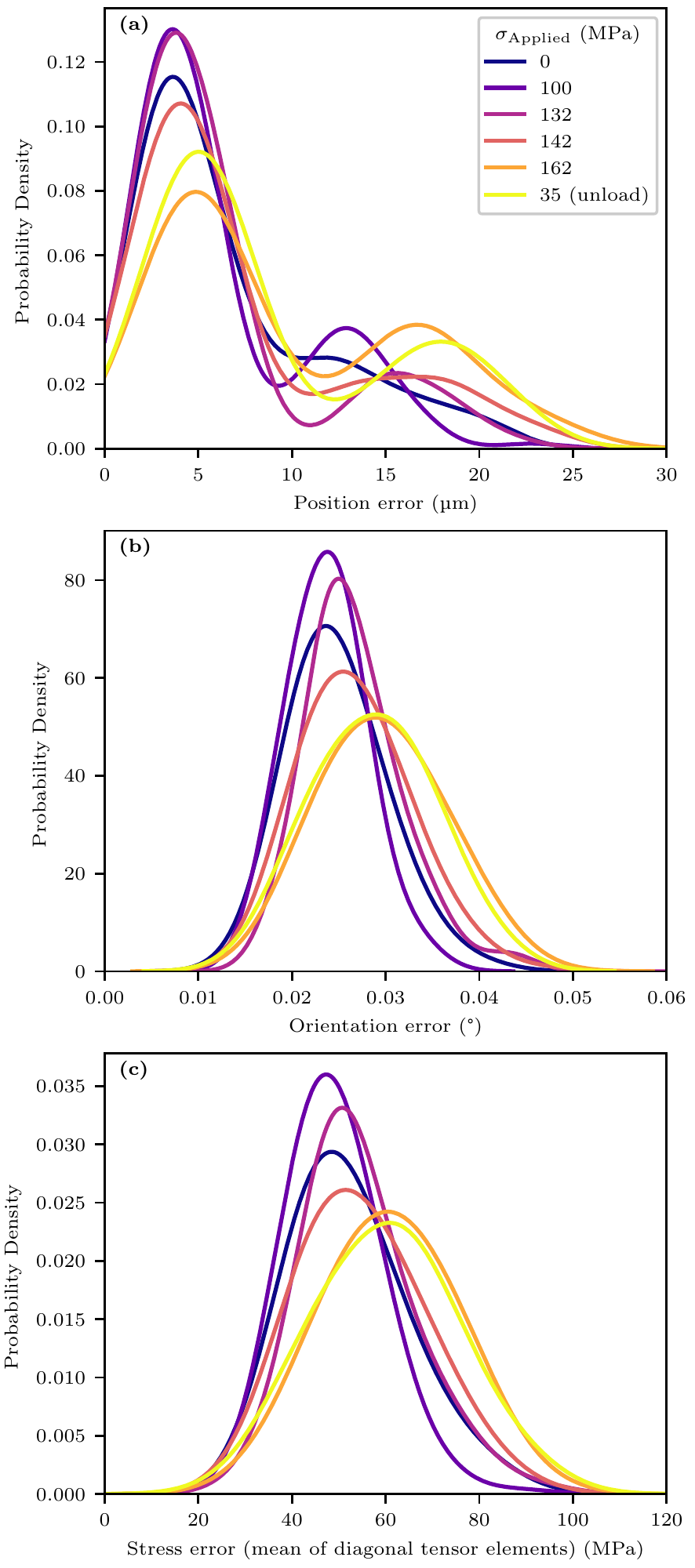}
    \caption{Per-grain error parameter distributions, shown for each loading step.}
    \label{fig:error_distributions}
\end{figure}

Using the bootstrap approach for grain error determination, Figure~\ref{fig:error_distributions} shows the distributions of errors in grain centre-of-mass position (a), orientation, (b) and stress tensor elements (c).
Grain volume-weighted averages of these distributions are provided in Table~\ref{tab:mean_errors}.
Position errors follow a broadly bimodal distribution, with most position errors around \SI{5}{\micro\metre}.
The outlying grains with larger position errors are likely caused by grains truncated by the X-ray beam at the top and bottom of the illuminated volume. As these values remain significantly below the average grain size, the conclusions of this work reliant on grain position remain valid.
Orientation and stress error histograms are normally distributed, and increase slightly with increasing applied load, as expected from the increased diffracted peak spread due to plastic mosaicity \cite{Poulsen2012}.
\begin{table}[H]
\centering
\caption{Mean grain parameter errors at each load step.}
\label{tab:mean_errors}
\begin{tabular}{@{}SSSS@{}}
\toprule
\begin{tabular}[c]{@{}c@{}}$\sigma_{\rm{Applied}}$\\(\si{\mega\pascal})\end{tabular} & \begin{tabular}[c]{@{}c@{}}Position\\(\SI{}{\micro\metre})\end{tabular} & \begin{tabular}[c]{@{}c@{}}Orientation\\ (\SI{}{\degree})\end{tabular} & \begin{tabular}[c]{@{}c@{}}Stress\\ (\SI{}{\mega\pascal})\end{tabular} \\ \midrule
0   & 3 & 0.02 & 62 \\
100 & 3 & 0.02 & 50 \\
132 & 4 & 0.03 & 53 \\
142 & 4 & 0.03 & 52 \\
162 & 5 & 0.03 & 58 \\
35  & 5 & 0.03 & 64 \\ \bottomrule
\end{tabular}
\end{table}

\subsection{Grain parameter correlations}

\begin{figure}[h]
    \includegraphics{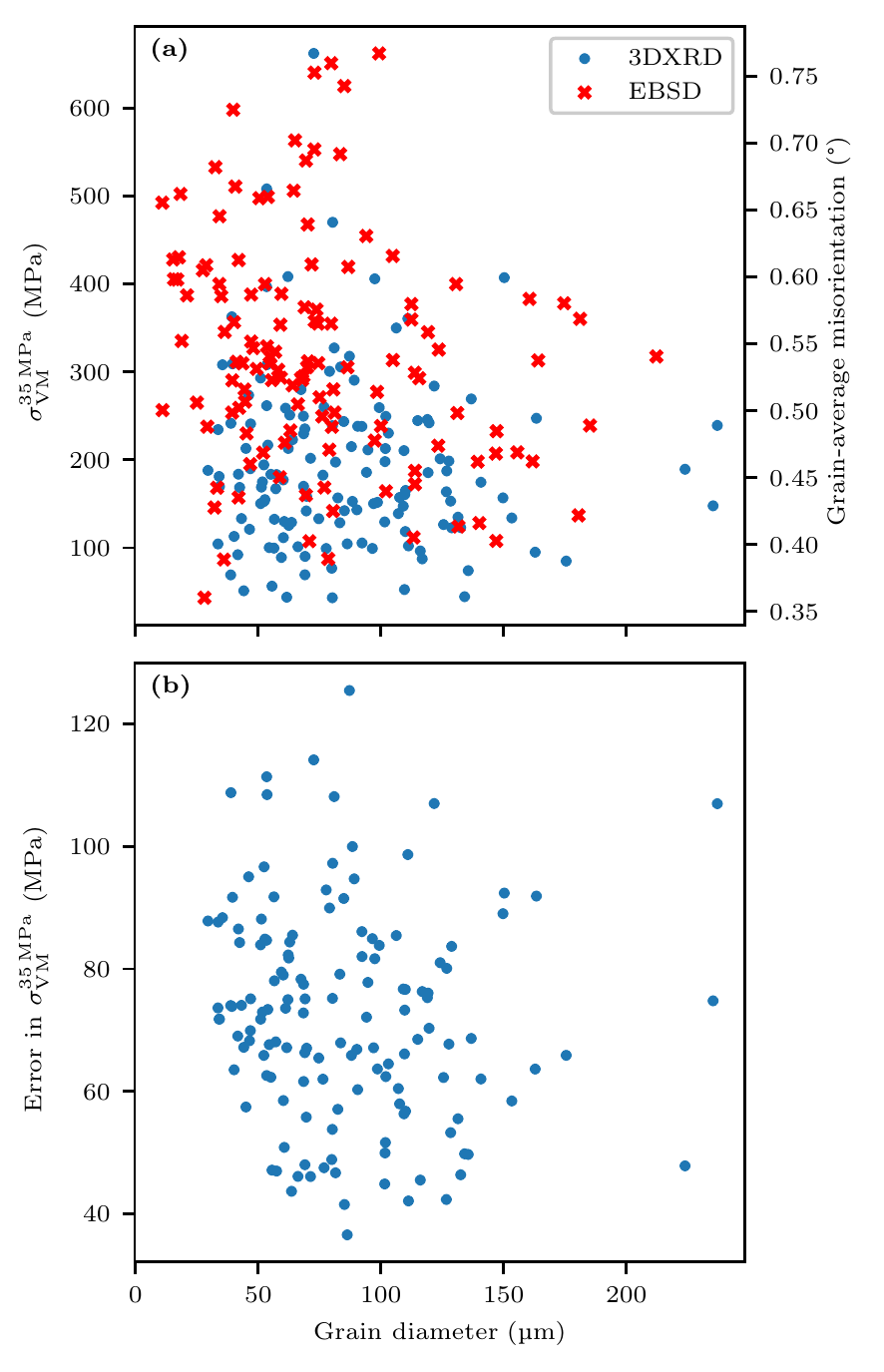}
        \caption{Grain diameter relationship in the deformed, unloaded condition ($\sigma_{\rm{Applied}} = \SI{35}{\mega\pascal}$) to the (a) 3DXRD grain von~Mises stress \& EBSD grain-averaged misorientation, and (b) the 3DXRD von~Mises stress error.}

    \label{fig:size_vs_vm_and_error}
\end{figure}

Grain volumes measured using far-field 3DXRD alone are determined using the mean intensity of the X-ray scattering peaks of that specific grain relative to the intensity of scattering peaks from all other grains \cite{nervo_comparison_2014}.
From these values, the grain size can be estimated by multiplying each 3DXRD grain volume by a constant scale factor.
Here, this corresponded to a value to match the grain size distribution obtained from EBSD measurements.
Comparing grain diameter against grain stress in the lab frame, as shown in Figure~\ref{fig:size_vs_vm_and_error}, yields an inverse correlation between the grain diameter and both von~Mises stress and error in von~Mises stress. 
Stresses were smaller and closer to the macroscopic applied stress for larger grains than smaller grains.
This behaviour is evidence of a Hall-Petch dependency at the intergranular level, which is heavily governed by intragranular grain-size dependent backstresses \cite{feaugas2003}.
To validate this as a real phenomenon, and not an artefact from the grain parameter refinement process, the grain-averaged misorientation from HR-EBSD is used as an analog for the 3DXRD intergranular stress.
As is evident in Figure~\ref{fig:size_vs_vm_and_error}, both follow a matching inverse relationship to the grain size.
The misorientation-grain size relationship is explained by the heterogeneous nature of plasticity; greater misorientation is correlated with increased residual elastic strains/stresses \cite{THOOL2020102785}.

\begin{figure}[h]
    \includegraphics[width=\columnwidth]{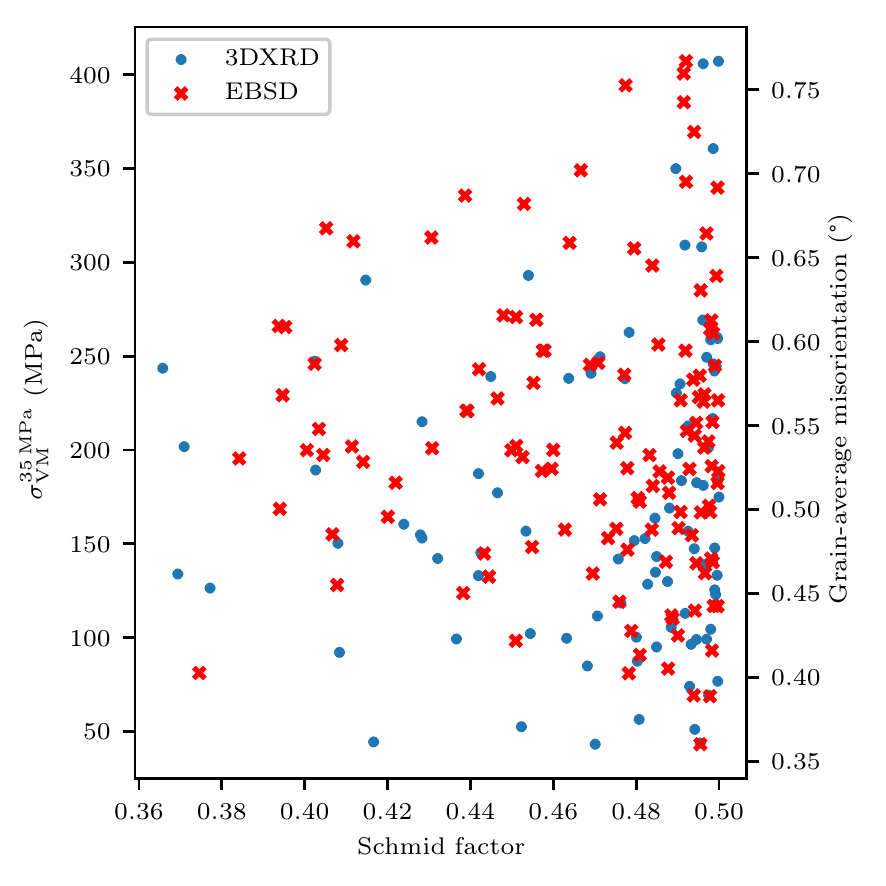}
    \caption{Tracked grain unloaded condition ($\sigma_{\rm{Applied}} = \SI{35}{\mega\pascal}$) von~Mises stress from 3DXRD data (blue dots) and EBSD grain-averaged misorientation (red crosses) vs grain Schmid factor.}
    \label{fig:vm_unload_ebsd_schmid}
\end{figure}

Again using the grain-averaged misorientation from EBSD data as an analogue for intergranular stress, the relationship between 3DXRD-measured grain stresses and the final Schmid factor is explored in Figure~\ref{fig:vm_unload_ebsd_schmid}. 
The relationship replicates the observation of the intergranular von~Mises stress change at yield, as in Figure~\ref{fig:combo_graph}c, where grains with a high Schmid factor have a larger allowed range of stresses, and grains with a lower Schmid factor are much more restricted in stress.
This is an intriguing observation as it implies that even though a grain may be favourably orientated for easy slip, it may experience elastic stresses that are (i) very high (behaving as a {\it{hard}} grain), (ii) moderate stresses where plasticity is easy, (behaving as a {\it{soft}} grain), or (iii) stresses below the critical resolved shear stress, so no slip at all.
The range of conditions indicates that grains with a high Schmid factor are more susceptible to grain-neighbour interactions, in a way that grains with a lower Schmid factor are not. 

Whilst intergranular stresses as measured via 3DXRD are clearly important, grain neighbour interactions from compatibility (amongst other mechanisms) can generate significant intragranular stress gradients \cite{BASU201711}.
Any stress localisation, governed by Type III stresses in particular for BCC steels, is predicted to ultimately determine locations of failure \cite{ERINOSHO2013170}.
Given the magnitude of the Type III stresses present in this material, significantly higher than the 3DXRD measured Type II stresses, their role cannot be neglected. The enormous difference between the Type II and Type III stress distribution widths were similarly observed by \citet{hayashi_intragranular_2019}, which also proposes that locations within grains of high triaxial stresses have low plastic strains, with adjacent regions compensating with large plastic strains.
Deformation in the present study is highly heterogeneous (see Figure~\ref{fig:hr_ebsd}b), with the behaviour of a given grain influenced by both Type II and Type III stress, its own and of the neighbours.

\subsection{Grain rotation during straining}

\begin{figure*}[h!]
    \centering
    \includegraphics{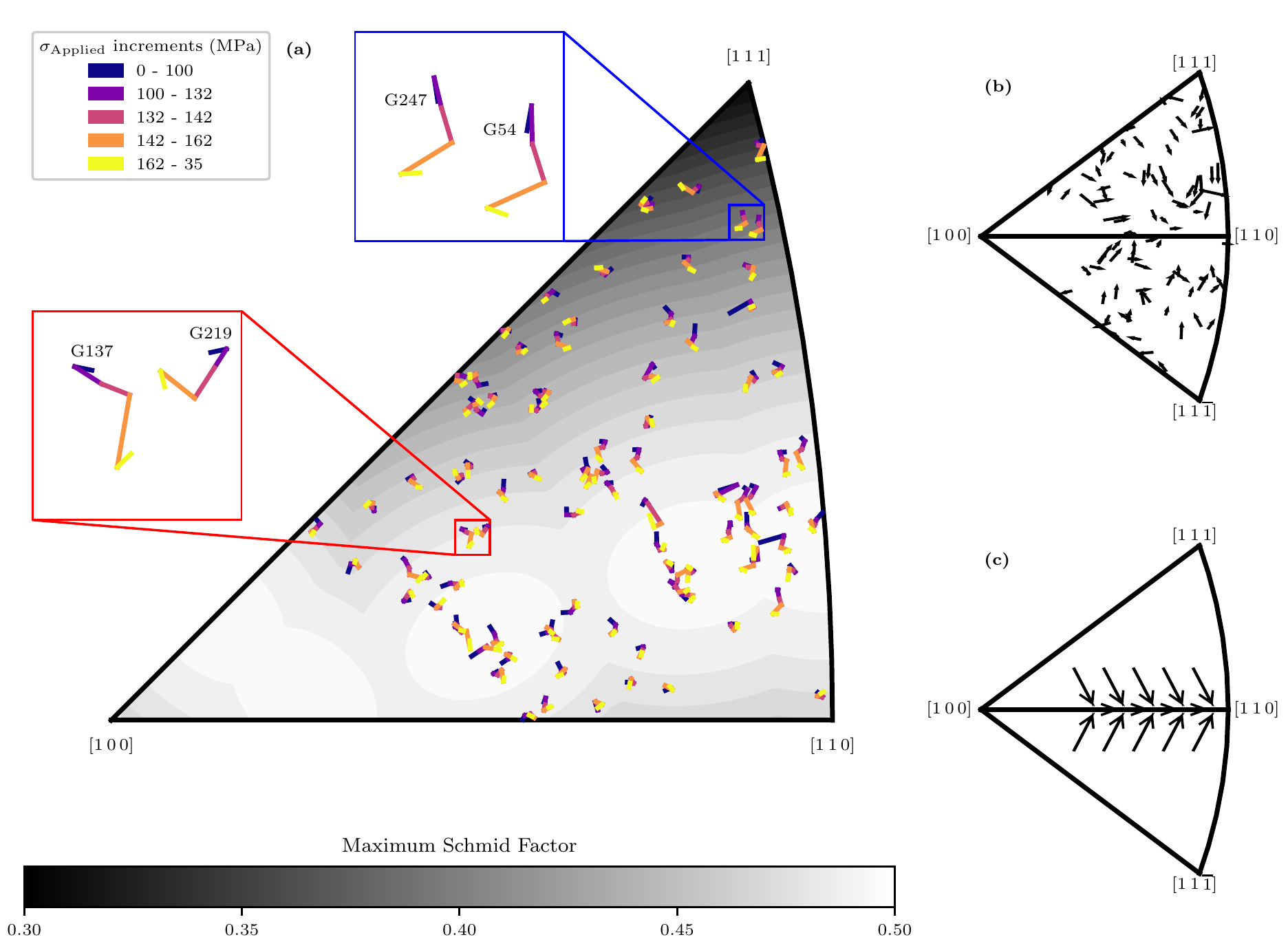}
    \caption{$Z$-axis inverse pole-figure, stereographic projection (shaded by Schmid factor), showing tracked grain rotations across multiple load steps (a), between no load ($\sigma_{\rm{Applied}} = \SI{0}{\mega\pascal}$) and max load ($\sigma_{\rm{Applied}} = \SI{162}{\mega\pascal}$) (b), and predicted rotations of BCC grains deforming by \hkl<111> pencil glide \citep{hosford_mechanical_2010} (c).}
    \label{fig:ipf_rotation}
\end{figure*}

Given there was strong evidence that the orientation (i.e. Schmid factor) will influence the stress development, the orientation changes of individual grains are considered.
Figure~\ref{fig:ipf_rotation}a shows the evolution in tracked grain orientation, plotted on an inverse pole figure, over all load steps.
Most grain rotations are seen to be small in magnitude.
This may be expected give the total plastic strain is \SI{\sim 5}{\percent}.
Interestingly, grain transformations are markedly different across load steps, with large changes in orientation direction visible once the applied stress passes the macroscopic yield point (at $\sigma_{\rm{Applied}} = \SI{100}{\mega\pascal}$). 
For certain similarly-orientated grains, such as those labelled G54 \& G247, the grain orientation changes are very similar in direction, indicating a strong correlation between grain initial orientation and subsequent rotation transformation under load. 
Other similarly-orientated grains, such as grains G137 and G219, have markedly different orientation changes under load.
One key difference between these two pairs of grains is their Schmid factor, with values for G54 \& G247 significantly lower than G137 \& G219.
This may be the reason a narrower stress range was observed for grains with a lower Schmid factor (see Figure~\ref{fig:combo_graph}c), and a wide range of stress states for those with a high Schmid factor, affecting their magnitude of rotation.
Whilst the observations here are limited to a few grains, observations of grains with similar initial orientations but dramatically different stress development were reported by \citet{hedstrom_load_2010} in an in-situ investigation.
This also reflects the findings of \citet{juul_analysis_2020}, who found that in-situ grain rotations under load were much more scattered than predicted by finite-element models.
Figure~\ref{fig:ipf_rotation}b further explores this discrepancy by plotting the overall grain orientation to the maximum applied load---the majority of tracked grains are observed to rotate as predicted for BCC crystals deforming by \hkl<111> pencil glide \citep{hosford_mechanical_2010} as per Figure~\ref{fig:ipf_rotation}c, although some grains do not rotate as expected from their starting 
orientation.
This is primarily because the rotation of a grain, at any single moment in time, is governed by its intergranular and intragranular stress state. 
It is evident from this study that the global Schmid factor alone cannot be used to predict the stress state for a given grain; it must instead be controlled by grain interactions.

\subsection{Grain neighbourhood}

There are several factors that are known to contribute to the stress state of the grain in polycrystals, which are often well correlated to significant orientation gradients.
Firstly, this depends on the initial orientation of a strained crystal and its orientation path as plasticity develops \cite{Raabe1994,LIU19985819,buchheit_carroll_clark_boyce_2015}.
The magnitude of per-grain stresses may well develop from any remnant residual elastic stress, accumulating from low strains, or present from prior processing \cite{HUGHES19973871, AGIUS2022103249}; this was unequivocally evident in the present study for grains with a measured stress initial stress inversely proportional to the subsequently accumulated stress change (see Figure~\ref{fig:combo_graph}b and \ref{fig:combo_graph}d).
Within a grain, there may be domains or differing orientations that gives rise to different dislocation glide systems, and thereby hardening \cite{HUGHES1997105, HUGHES19973871,raabe_theory_2002}.
Continued subdivision of cell structures developing within a grain, arising from plasticity structures, will themselves influence both the inter- and intragranular stresses \cite{BECKER19952701, HUGHES19973871}.
Finally, the grain stress will also depend on the stress and crystallography attributes of the neighbouring microstructure environment \cite{kocks_texture_2000, raabe_theory_2002, bretin_neighborhood_2019, abdolvand_strong_2018}.

\begin{figure*}[!ht]
    \centering
    \includegraphics{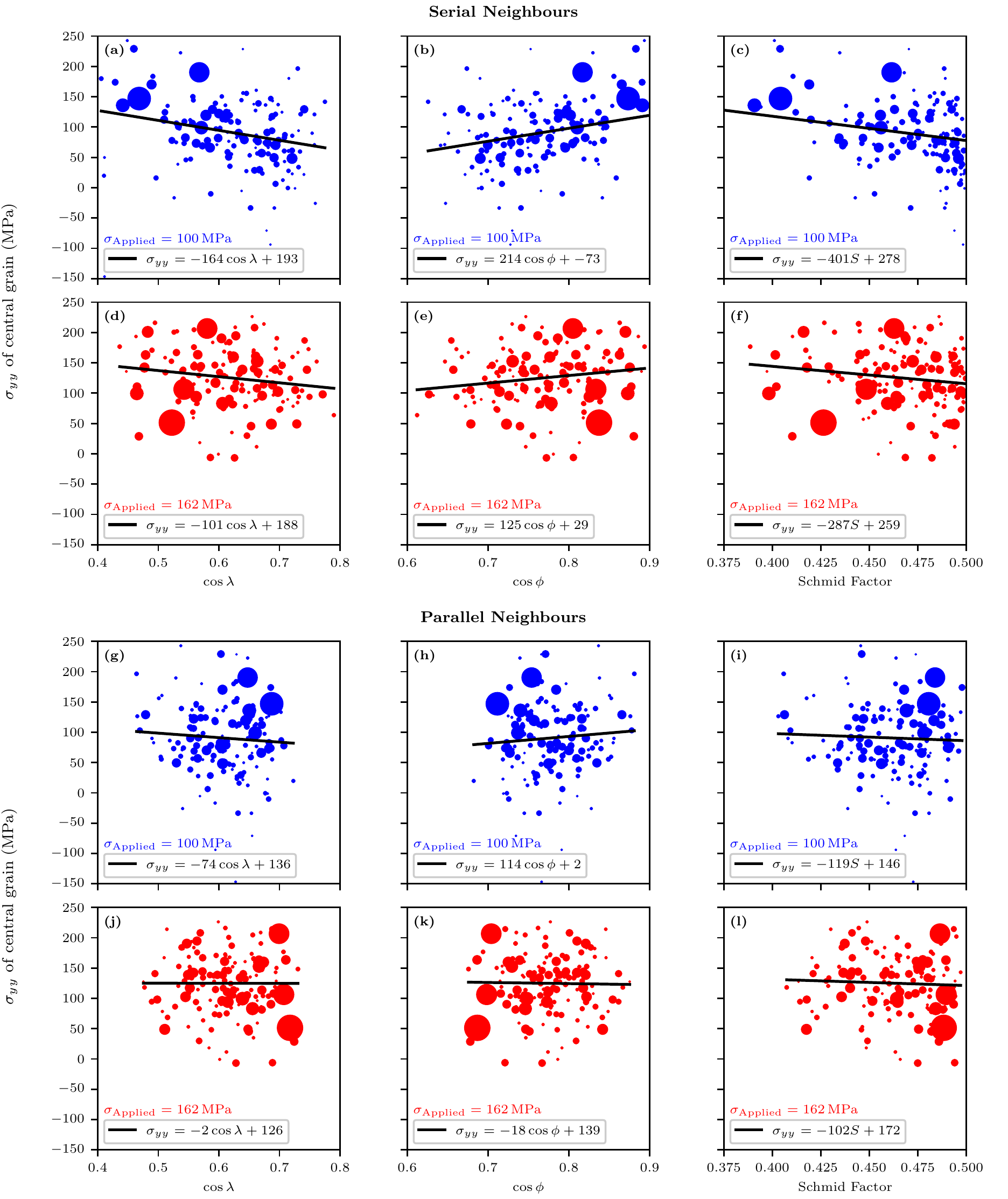}
    \caption{Dependence of grain stress in the tensile direction, $\sigma_{yy}$, on the slip direction component, $\cos\lambda$, the slip plane normal component, $\cos{\phi}$, and the Schmid factor. This is shown for both serial and parallel neighbours, and for each at the onset of plasticity, $\sigma_{\rm Applied} = \SI{100}{\mega\pascal}$ and at \SI{\sim 5}{\percent} plastic strain, $\sigma_{\rm Applied} = \SI{162}{\mega\pascal}$. The size of the circles represent the relative volumes of the grains.}
    \label{fig:neigh_oriens}
\end{figure*}

For a given grain of interest, the influence of neighbouring grains is explored here. 
For each grain, a Schmid factor was devised by taking the highest Schmid factor for slip in any \hkl<111> direction.
Here, a {\it{soft}} grain is defined as a crystal exhibiting a high Schmid factor that is well aligned for easy slip, relieving stress, whereas a {\it{hard}} grain is orientated poorly for easy glide and would be expected to build higher elastic stresses.
The existence of such effects are well reported in alloys with HCP crystal structures \cite{DUNNE20071061}, but its significance is seldom reported for highly symmetric cubic systems.
To ascertain the effect in ferritic steel, the nearest neighbours of each grain using a Delaunay triangulation \cite{2020SciPy-NMeth} were identified at (i) the yield point, $\sigma_{\rm{Applied}} = \SI{100}{\mega\pascal}$, and (ii) at \SI{\sim 5}{\percent} plastic strain, $\sigma_{\rm{Applied}} = \SI{162}{\mega\pascal}$. 
Neighbours were then subdivided into series and parallel neighbours, depending on their position relative to the central grain. 
From a vector that connects grain-to-grain centroid coordinates, an inclination angle was calculated between this and the tensile axis. Serial neighbours were classified as those with an inclination angle less than \SI{45}{\degree} and parallel neighbours were those with an inclination angle greater than \SI{45}{\degree}; such classification has been established by other authors \cite{abdolvand_strong_2018}. 
Whether a grain is {\it{hard}} or {\it{soft}} could next be ascertained for serial and parallel neighbours of a central grain of interest, calculated for the tensile direction stress, $\sigma_{yy}$.
This analysis is reported in Figure~\ref{fig:neigh_oriens}; stresses of a central grain are provided as a function of the average Schmid factor of its serial and parallel neighbours, but also for its components $\cos\lambda$ and $\cos\phi$ to explore which of these components govern the response, as defined by the Schmid factor equation for the critically resolved shear stress, $\tau$:
\begin{equation}
    \tau = \cos(\phi)\cos(\lambda)\sigma
\end{equation}
where $\phi$ is the angle between the slip plane normal and the tensile axis, $\lambda$ is the angle between the slip direction and the tensile axis, and $\sigma$ is the applied stress.
While the scatter is large, a negative correlation is evident between the mean Schmid factor of serial neighbours and the stress achieved by that grain, Figure~\ref{fig:neigh_oriens}c at the onset of yield ($\sigma_{\rm{Applied}} = \SI{100}{\mega\pascal}$). 
If a grain is in series with {\it{hard}} grains (low Schmid factor), it tends to experience a greater stress in the loading direction compared to softer grains (high Schmid factor). 
This finding, reported in-situ for a BCC system for the first time, replicates the trend for HCP crystals \cite{abdolvand_strong_2018}, though critically, is a weaker effect.
By plotting the individual components of the Schmid factor, the influence of slip direction vs slip plane normal orientation relative to the loading axis can be separately observed.
For serial neighbours, the negative correlation between a central grain $\sigma_{yy}$ and the Schmid factor is also observed with $\cos{\lambda}$; interestingly, this is the converse to the slip plane normal component behaviour, $\cos\phi$ (Figure~\ref{fig:neigh_oriens}b), showing a positive correlation with $\sigma_{yy}$ stress.
This indicates that the slip direction in an adjoining grain determines the Schmid factor dependent neighbourhood effect.
The analysis is also shown for $\sigma_{\rm{Applied}} = \SI{162}{\mega\pascal}$ (Figure~\ref{fig:neigh_oriens}d--f).
The trends replicate the observations at the yield stress, but the gradients are shallower.
This indicates that the neighbourhood effect diminishes as plastic strain increases.
One can postulate that such neighborhood effects may disappear entirely as plasticity builds; grains must continue to rotate, which is likely to eliminate any grain-to-grain load partitioning as their respective Schmid factors become more similar.
Given there are 48 slip systems for BCC, with a great degree of freedom for crystal distortion/rotation, this operation will be easy.
It is also plausible that the elimination of grain neighbour stress partitioning, which would otherwise be detrimental to fracture strain, is a key reason why these materials exhibit exceptional ductility.

The effect of parallel neighbours on central grain maximum stress is minimal in this case (see Figure~\ref{fig:neigh_oriens}g--l), matching observations by \citet{abdolvand_strong_2018}. There is no strong relationship between the mean Schmid factor of neighbouring grains, and the central grain $\sigma_{yy}$ stress; the trend lines when at $\sigma_{\rm{Applied}} = \SI{162}{\mega\pascal}$ (Figure~\ref{fig:neigh_oriens}j--l) are notably flat.
In short, the axial stress for a central grain is correlated with the orientation of neighbouring grains located in series along the loading axis with it. 
Neighbouring serial grains with a lower Schmid factor lead to higher overall axial stresses in the central grain.
This supports the conclusions of prior modelling studies of cubic systems by \citet{bretin_neighborhood_2019}, who found that the influence of neighbouring grain orientations on the stress state depends on the relative position of the central grain and the neighbouring grain with respect to the loading axis.
This observation has implications for future modelling endeavours---for crystal plasticity finite element method (CPFEM) simulations of a very small number of grains, for example, simply randomising all grain orientations may not be sufficient to remove influences of nearest neighbour orientation on grain stress states.
Instead, a number of model iterations with shuffled grain orientations may be required, as observed in face-centered cubic systems by \citet{kocks_texture_2000}.
Na\"{i}ve simulations (with a larger number of grains) of macroscopic stress anisotropy due to sample texture may be similarly affected by ``unlucky" shuffles of grain orientation due to this effect.
To establish the significance of the grain neighbour effect observed in the present study, on a cubic system, future investigations at high plastic strains, for other cubic structure polycrystals, and the role of Type III intragranular stresses are exciting future areas for exploration in this field.

\clearpage

\clearpage

\section{Conclusion}

An in-situ three-dimensional X-ray diffraction experiment during mechanical loading of a low-carbon ferritic steel has been performed to reveal individual crystal behaviour. Significant developments on existing 3DXRD analysis methods have revealed several insights, which are summarised here:
\begin{enumerate}
    \item{This experiment is the first example of an in-situ 3DXRD experiment at Diamond Light Source, demonstrating the feasibility to track the response of a polycrystalline material during deformation on a per-grain level.}
    
    \item{A rigorous approach to quantifying uncertainties related to per grain stress and orientation analysis provide confidence that reliable in-situ 3DXRD measurements can be performed on I12 at Diamond. Confidence in orientation is stable at around \SI{0.03}{\degree}, and the stress error varies between \SIrange{50}{64}{\mega\pascal}}
 
    \item{The initial Schmid factor of a tracked grain was found to influence the increase in the per-grain von~Mises stress. Grains with a low initial Schmid factor were found to almost exclusively increase in stress, whereas grains with a high Schmid factor had a wider range of allowed changes in stress.}
   
    \item{The change in per-grain von~Mises stresses between the first two load steps was found to be strongly negatively correlated to the initial von~Mises stress of a grain. This demonstrates that tracked grain residual stresses are dominant over subsequent grain stress evolution.}
    
    \item{Grain $\sigma_{yy}$ (axial) stresses were found to broadly follow the macroscopic stress-strain curve, but a significant fraction of grains experienced a stress drop beyond the global yield point. The magnitude of the stress drop was found to be proportional to the grain residual stress---grains with a higher starting $\sigma_{yy}$ tended to experience a greater stress drop.}
      
    \item{Most grains followed an orientation change that is expected from \hkl<111> pencil glide, however, examples that do not follow this trend were evident; similarly orientated grains may posses quite dissimilar orientation paths during straining. This may be explained by their Schmid factor, stress state, and grain neighbour interactions.}
    
    \item{A grain neighbourhood effect is evident at low plastic strains. Grains with {\it{hard}} series neighbours were found to have a higher axial stress at the macroscopic yield point, whilst parallel neighbours had limited effect on a grain stress state. The strength of this effect diminishes at higher macroscopic strains, as grains rotate and Schmid factors between neighbours become more similar, load partitioning becomes less significant. The ability for ferritic steels to eliminate these neighbour effects is believed to contribute to the exceptional ductility they possess.} 
\end{enumerate}

\section{Acknowledgments}

This work was supported by Diamond Light Source, instrument I12 [NT26376] and the Engineering and Physical Sciences Research Council [EP/R030537/1].
James Ball would like to thank the Diamond Light Source and the University of Birmingham for jointly funding his PhD program, as well as Anastasia Vrettou and Neal Parkes for their help with the sample preparation process, and Younes El-Hachi and Jon Wright for their assistance with the bootstrap method of grain parameter error determination.

\clearpage
\typeout{}
\bibliography{main}
\bibliographystyle{elsarticle-num-names}

\clearpage
\appendix
\section{Strain and stress tensor determination}
\label{sec:tensors}
The Biot strain is used as the default representation of lattice strain for grains defined by ImageD11.
The derivation of the Biot strain tensor from grain lattice lengths is provided here. 
First, the deformation gradient tensor $\tens{F}$ is determined:
\begin{equation}
\tens{F} = \left(\matr{U}\cdot\matr{B}\right)^{-\intercal}\cdot{\matr{B}_{0}^{\intercal}}
\end{equation}
where $\matr{U}$ rotates a vector in the Cartesian grain system, $\vec{G}_c$, to the sample reference frame, $\vec{G}_s$:
\begin{equation}
\vec{G}_s = \matr{U}\vec{G}_c
\end{equation}
$\matr{B}$ transforms a vector in reciprocal space, $\vec{G}_{hkl}$, to the Cartesian grain system $\vec{G}_c$ \citep{poulsen_three-dimensional_2004}:
\begin{equation}
\vec{G}_c = \matr{B}\vec{G}_{hkl}
\end{equation}
and is defined by the deformed lattice parameters of the grain in reciprocal space $\left(a^*, b^*, c^*, \alpha^*, \beta^*, \gamma^*\right)$:
\begin{equation}
\matr{B} = \begin{pmatrix}
a^* & b^*\cos\left(\gamma^*\right) & c^*\cos\left(\beta^*\right)\\
0 & b^*\sin\left(\gamma^*\right) & -c^*\sin\left(\beta^*\right)\cos\left(\alpha\right)\\
0 & 0 & c^*\sin\left(\beta^*\right)\sin\left(\alpha\right)
\end{pmatrix}
\end{equation}
and:
\begin{equation}
\cos\left(\alpha\right) = \frac{\cos\left(\beta^*\right)\cos\left(\gamma^*\right)-\cos\left(\alpha^*\right)}{\sin\left(\beta^*\right)\sin\left(\gamma^*\right)}
\end{equation}
$\matr{B}_{0}$ is calculated from the reference unit cell lattice lengths.
$\tens{F}$ is then used to determine the Biot strain tensor $\tens{E}$:
\begin{equation}
\tens{E} = \tens{C}^{\nicefrac{1}{2}} - \matr{I} = \left(\tens{F}^{\intercal}\cdot\tens{F}\right)^{\nicefrac{1}{2}}-\matr{I}
\end{equation}
where $\tens{C}$ is the right Cauchy-Green deformation tensor and $\matr{I}$ is the identity matrix. 
ImageD11 uses polar composition to determine $\tens{C}^{\nicefrac{1}{2}}$:
\begin{equation}
\tens{F} = \matr{V}\cdot\matr{R} = \matr{R}\cdot\tens{C}^{\nicefrac{1}{2}}
\end{equation}
The stress tensor $\tens{\sigma}$ is then determined from the strain tensor as defined by \citet{oddershede_determining_2010} using the stiffness values in Table~\ref{tab:stiffness}.
\begin{table}[H]
\centering
\caption{Stiffness constants used for ferrite phase \citep{inal_second-order_2004}.}
\label{tab:stiffness}
\begin{tabular}{@{}lS@{}}
\toprule
Stiffness Constant & \multicolumn{1}{c}{Value (\SI{}{\giga\pascal})} \\ \midrule
$c_{11}$           & 231.4       \\
$c_{12}$           & 134.7       \\
$c_{44}$           & 116.4       \\ \bottomrule
\end{tabular}
\end{table}
It is noted that the Biot strain tensor $\tens{E}$ determined by ImageD11 is not equivalent to the linear Lagrangian strain tensor $\tens{\varepsilon}$ used by \citet{oddershede_determining_2010}. 
Therefore, the Biot stress tensor equivalent does differ from $\tens{\sigma}$.
However, the tensors must be equivalent in the small strain limit, and the error between representations is assumed to be negligible at the strain levels observed in this study. 
The strain and stress tensors in the sample coordinate system ($\tens{E}_{s}$ and $\tens{\sigma}_{s}$ respectively) are also determined by rotating $\tens{E}$ and $\tens{\sigma}$ \citep{oddershede_determining_2010}:
\begin{align}
\tens{E}_{s} &= \matr{U}\tens{E}\matr{U}^\intercal\\
\tens{\sigma}_{s} &= \matr{U}\tens{\sigma}\matr{U}^\intercal
\end{align}

\section{Grain de-duplication algorithm}
\label{sec:dedup}
A new de-duplication algorithm has been devised to identify repeated observations of the same grain, given a list of grains, a distance tolerance, and a misorientation tolerance. 
First, a list of all possible grain pairs within a specified centre-of-mass distance (the distance tolerance) is produced. 
A nearest-neighbour search algorithm using K-dimensional trees (as originally defined by \citet{maneewongvatana_analysis_1999}) is performed using an implementation in the scipy Python library \citep{2020SciPy-NMeth} to generate a list of candidate grain pairs. 
Once the candidate grain pairs list is identified, the misorientation between the grains of each pair is calculated. 
The misorientation algorithm, originally defined by \citet{proudhon_pymicro_2021} in the pymicro Python library, has been vectorised with the help of the Numba Python library \citep{lam_numba_2015}, then parallelised. 
This is much faster than the single-threaded approach---with a modern 12-core AMD Ryzen processor, calculating the misorientation of \num{100000} grain pairs takes \SI{5}{\second} with the parallelised approach, vs \SI{47}{\second} with the original approach.
Once the misorientation for each grain pair is determined, grain pairs with a misorientation that is greater than the misorientation tolerance are removed. 
The remaining grain pairs are therefore close together in position and orientation. 

Once the final grain pairs list is produced, a graph is constructed using the NetworkX Python library \citep{hagberg_exploring_2008}. 
Each node in the graph represents a grain, and nodes are connected with edges if the grains in the nodes are paired. 
Then, a list of all connected component sub-graphs is generated.
Each connected component sub-graph represents a group of nodes that is connected together.
For example, if grains \textbf{A} and \textbf{B} are paired, and grains \textbf{B} and \textbf{C} are paired, grains \textbf{A}, \textbf{B} and \textbf{C} form a connected component subgraph. 
This connected component sub-graph \textbf{ABC} is then said to represent a single physical grain, observed more than once.

In practice, the outputs of this algorithm are used slightly differently depending on the purpose. 
At the grain cleaning stage, all grains provided in the grain list are from the same individual 3DXRD scan.
Therefore, each grain group returned from the algorithm is associated to a single ``clean grain", which may consist of multiple raw observations, or ``raw grains". 
The parameters of the ``clean grain" (such as grain position, orientation, and strain state) are determined from the parameters of the raw grains. 
A volume-weighted average of raw grain positions is used for the position of the clean grain.
The $\matr{UBI}$ matrix of the clean grain (where $\matr{UBI} = \left(\matr{U}\cdot{\matr{B}}\right)^{-1}$) is taken from the $\matr{UBI}$ of the largest contributing raw grain.
The volume of the clean grain is determined by the sum of the volumes of the raw grains.
The same procedure applies to the letterbox stitching stage.
When the algorithm is applied to grains across load steps, multiple grains from each load step may be found in each grain group returned by the algorithm.
In this instance, only the largest grain in each load step is used.

\end{document}


\title{{\bf Per-grain and neighbourhood stress interactions during deformation of a ferritic steel obtained using three-dimensional X-ray diffraction}}

\author[a, b]{James A. D. Ball}
\author[c]{Anna Kareer}
\author[b]{Oxana V. Magdysyuk}
\author[b]{Stefan Michalik}
\author[b]{Thomas Connolley}
\author[a,*]{David M. Collins}
\affil[a]{{\small School of Metallurgy and Materials, University of Birmingham, Edgbaston, Birmingham, B15~2TT, United Kingdom}}
\affil[b]{{\small Diamond Light Source Ltd., Harwell Science and Innovation Campus, Didcot, OX11 0DE, United~Kingdom}}
\affil[c]{{\small Department of Materials, University of Oxford, Parks Road, Oxford, OX1 3PH, United Kingdom}}
\affil[*]{{\small Corresponding author: D.M.Collins, D.M.Collins@bham.ac.uk}}

\date{}
\maketitle

\vspace{-10pt}

\begin{center}
    {\Large\bf{Supplementary Content}}
\end{center}

\vspace{-10pt}

\noindent\makebox[\linewidth]{\rule{\textwidth}{0.4pt}}

\vspace{10pt}

\begin{figure}[h]
    \centering
    \includegraphics{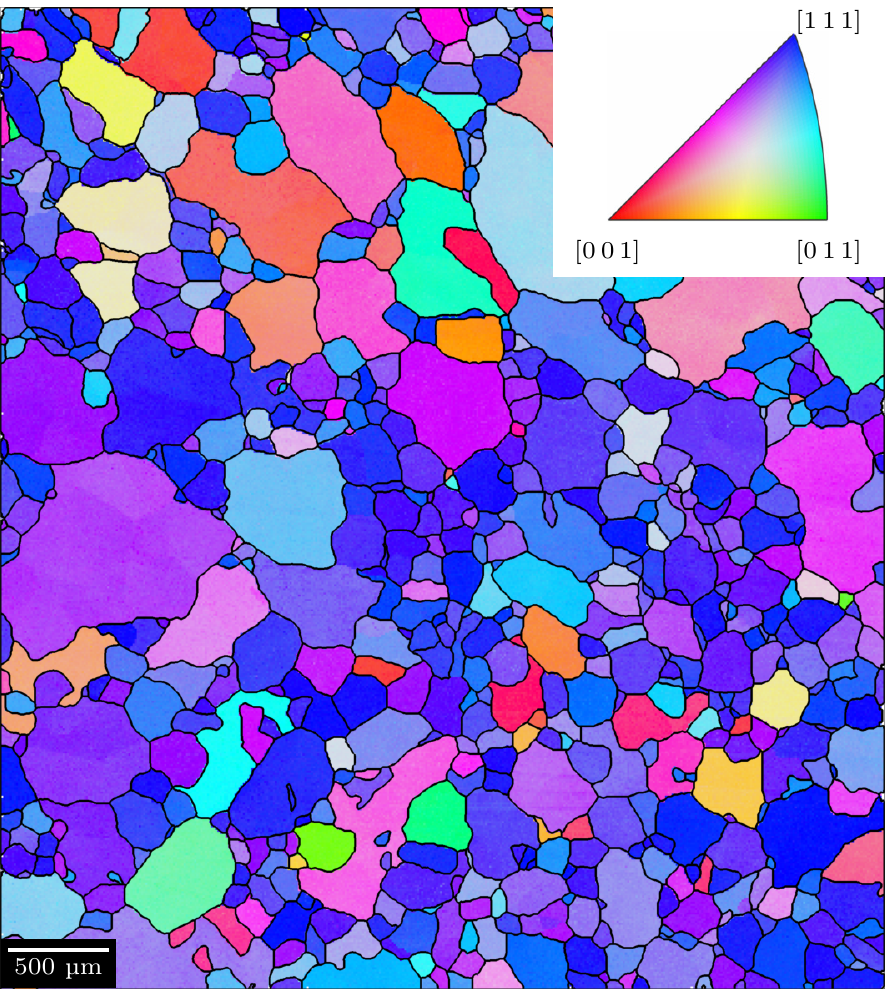}
    \caption{Low magnification IPF-Z map from EBSD data on undeformed DX54 ferritic steel.}
    \label{fig:EBSD_map}
\end{figure}

\noindent Figure~\ref{fig:EBSD_map} is a low magnification EBSD map from the undeformed grip section of the DX54 ferritic steel sample.
A large equiaxed single-phase ferritic microstructure is observed.
These data were used to determine the average grain size of the material.

\begin{figure}[h]
    \centering
    \includegraphics{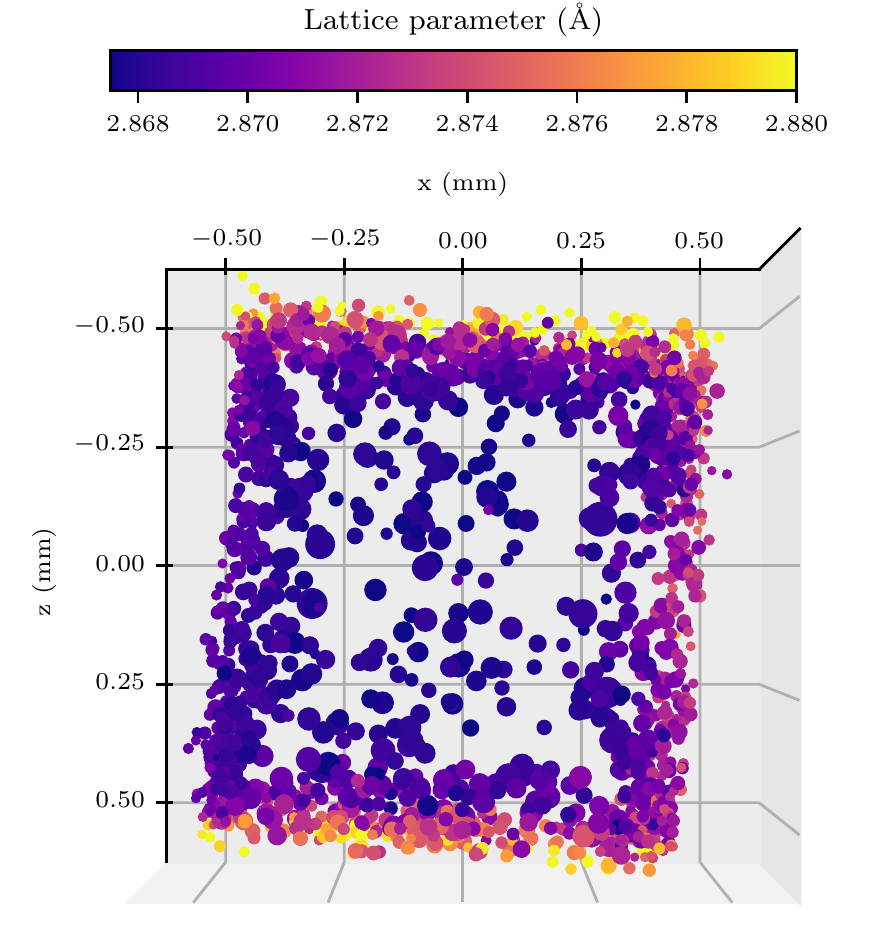}
    \caption{Grain map at no applied load, observed as a cross section of the dog bone specimen. The points here represent grains, which are coloured by lattice parameter and their size is scaled by their volume.}
    \label{fig:no_load_map_lattice}
\end{figure}

Figure~\ref{fig:no_load_map_lattice} is a map of 3DXRD grain centre-of-mass positions for the first load step ($\sigma_{\rm{Applied}} = \SI{0}{\mega\pascal}$).
Grain points in the figure are coloured by the mean crystal unit cell length of the grain, as determined from the $\matr{B}$ matrix by ImageD11.
A significant near-surface variation in grain lattice parameter is observed at the surfaces perpendicular to the $z$-axis.
These were the original external galvanized surfaces of the steel sheet from which the samples were cut.
Note the effect was not observed on sample surfaces perpendicular to the x-axis, which were originally in the bulk of the sheet.

\begin{figure}[h]
    \centering
     \includegraphics{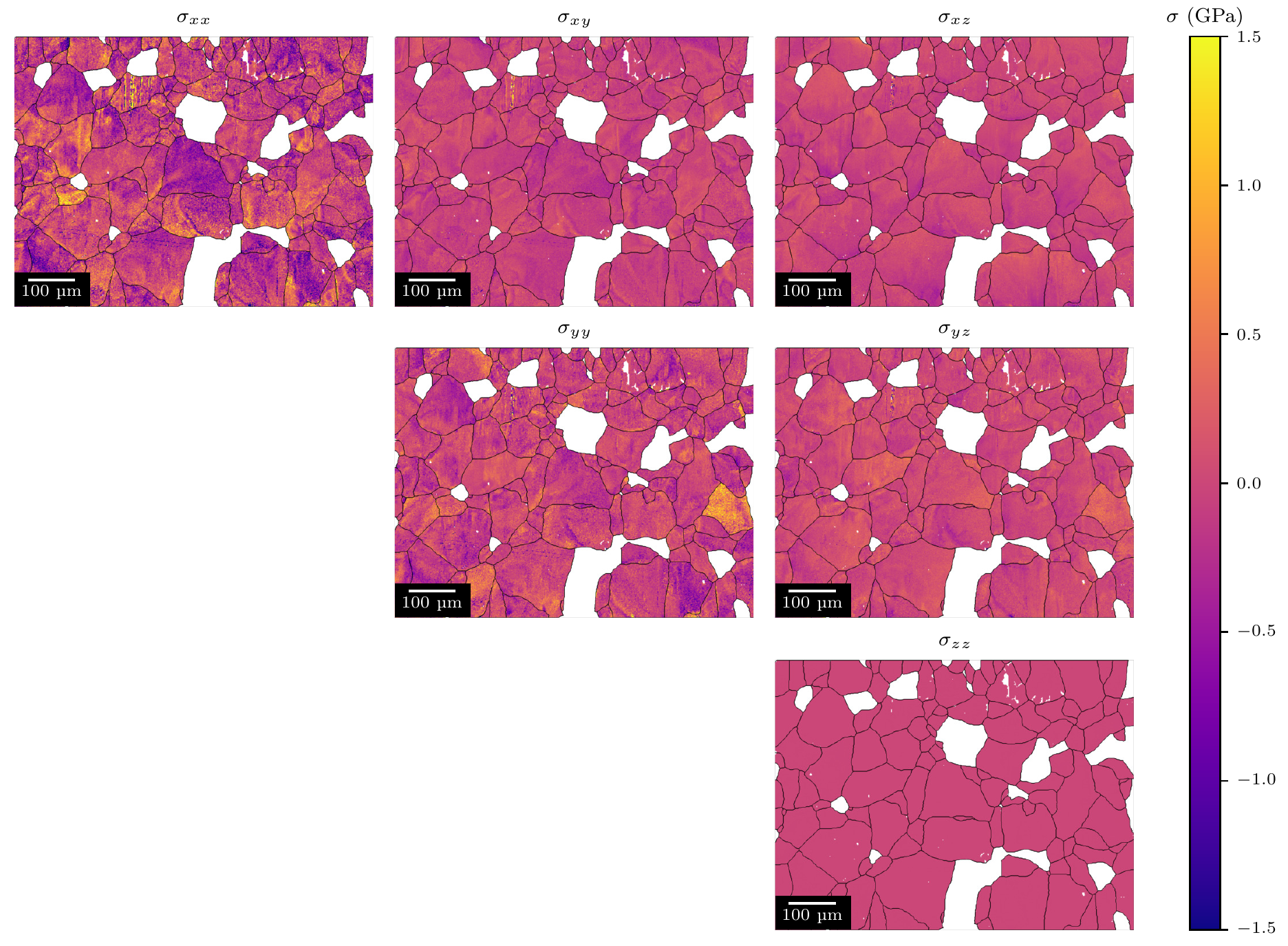}
    \caption{HR-EBSD maps obtained from the sample gauge, post deformation with \SI{\sim 5}{\percent} plastic strain and per-pixel Type III stress tensor elements. The loading axis is left-right.}
    \label{fig:ebsd_stress_grid}
\end{figure}

Figure~\ref{fig:ebsd_stress_grid} is a map of each component of the $\tens{\sigma}$ stress tensor obtained from the processed HR-EBSD data.
Outlying grains with non-physical stress distributions (due to poor quality data for these grains) have been masked in white.
Significantly wide stress distributions (on the order of \SI{\sim 3}{\giga\pascal}) are observed in the in-plane axial and transverse directions ($\sigma_{xx}$ and $\sigma_{yy}$ respectively).
Narrower distributions are observed in the shear directions ($\sigma_{xy}$, $\sigma_{xz}$ and $\sigma_{yz}$).

